\documentclass[preprint, authoryear, 5p, number]{elsarticle}
\pdfoutput=1
\usepackage{units}
\usepackage{amsmath}
\usepackage{amssymb}
\usepackage[english]{babel}
\usepackage{color}
\usepackage{array}
\usepackage{booktabs}
\usepackage{graphicx}
\usepackage{array}
\usepackage{gensymb}
\usepackage{relsize}
\usepackage[latin1]{inputenc}
\usepackage{float}
\usepackage{hyperref}
\usepackage{lscape}
\usepackage{csquotes}

\newcolumntype{M}[1]{>{\centering\arraybackslash}m{#1}}
\newcolumntype{N}{@{}m{0pt}@{}}
\newcommand*\Laplace{\mathop{}\!\mathbin\bigtriangleup}

\journal{Astronomy and Computing}
 
\begin{document}
 
\begin{frontmatter}
 
\title{Finding faint H\,{\large I} structure in and around galaxies: scraping the barrel}

 \author[a]{D.~Punzo\corref{cor1}}
 \ead{D.Punzo@astro.rug.nl}
 
 \author[a]{J.M.~van der Hulst}
 \author[b]{J.B.T.M.~Roerdink}

 \cortext[cor1]{Corresponding author}

 \address[a]{Kapteyn Astronomical Institute, University of Groningen, Landleven 12, 9747 AD Groningen, The Netherlands}
  
 \address[b]{Johann Bernoulli Institute for Mathematics and Computer Science, University of Groningen, Nijenborgh 9, 9747 AG Groningen, The Netherlands}
 
\begin{abstract}
Soon to be operational H\,{\small I} survey instruments such as APERTIF
and ASKAP will produce large datasets. 
These surveys will provide information about the H\,{\small I} in and around 
hundreds of galaxies with a typical 
signal-to-noise ratio of $\sim$ 10 in the 
inner regions and $\sim$ 1 in the outer regions. In addition, such surveys will make it 
possible to probe faint H\,{\small I} structures, typically located
in the vicinity of galaxies, such as extra-planar-gas, tails and filaments.
These structures are crucial for understanding galaxy evolution, particularly when they 
are studied in relation to the local environment.
Our aim is to find optimized kernels for the discovery of
faint and morphologically complex 
H\,{\small I} structures.
Therefore, using H\,{\small I} data from a variety of galaxies, we explore state-of-the-art
filtering algorithms. We show that the intensity-driven gradient filter, due to 
its adaptive characteristics, is the optimal choice. In fact, this filter requires 
only minimal tuning of the input parameters to enhance the signal-to-noise ratio of faint
components. In addition, it does not degrade 
the resolution of the high signal-to-noise component of a source.
The filtering process must be fast and be embedded in an interactive visualization
tool in order to support fast inspection of a large number of sources.
To achieve such interactive exploration, we implemented a multi-core CPU (OpenMP) 
and  a GPU (OpenGL) version of this filter in a 3D visualization environment ($\tt{SlicerAstro}$).

\end{abstract}
 
 
\end{frontmatter}

\section{Introduction}\label{intro}
Radio data are intrinsically noisy and most sources are faint and often extended 
(see for example the WHISP catalog, \cite{Whisp}). Very faint coherent H\,{\small I} signals,
below a 3 sigma \textit{rms} noise level, are difficult to find \citep{Popping}. Depending on the source 
structure, spatial and/or spectral smoothing can increase the signal-to-noise ratio.
Smoothing is usually applied to multiple spatial and spectral scales to ensure that sources 
of different size are extracted at their maximum integrated signal-to-noise ratio.

In upcoming blind H\,{\small I} surveys such as WALLABY, using the ASKAP telescope \citep{askap, duffy}, 
and the shallow and medium-deep APERTIF surveys, using the WSRT telescope \citep{Apertif3}, 
source finding will be a major concern. Source finders \citep[e.g.,][]{Whiting, sofia} are 
designed to automatically detect all the sources in the field and to achieve this goal they
must employ an efficient mechanism to discriminate between interesting candidate sources and 
noise. Due to the 
complex 3-D nature of the sources \citep{Sancisi} and the noisy character of the data,
constructing a fully automated and reliable pipeline is not trivial. \citet{Popping} reviewed 
the current state of the art and described the issues connected with 
the noisy nature of the data, and the various methods and their efficiency.  

In the source-finding process, \textit{masks} are generated enclosing the sources. 
The determination of the final masks involves a variety of filtering 
operations in order to pick up faint and extended emissions.  
However, users are ultimately provided with
the mask and data products determined from the original data within the 
masks. In order to examine the original data within and around
the mask, to check the performance of the source 
finding process and to investigate whether all faint structures have been included, it 
is necessary to have a visualization tool that not only shows the original data and 
the mask, but also has the ability to interactively filter 
the data to bring out the very faint structures in the data.

Our goal therefore is the development of a suitable filtering method in a 3D 
visualization environment that maximizes the local signal-to-noise ratio 
of the very faint structures (signal-to-noise ratio $\sim 1$)
while preserving its specific 3-D structure (e.g. tidal tails,
filaments and extra-planar gas). Ideally, the method should be 
adaptive (in such a way that the user
does not have to explore a large parameter space to get the best result), interactive, and 
fast, i.e. applicable in real-time. In this paper,
we explore a number of existing filtering methods in combination with a 3D 
visualization tool \citep{Punzo} in order to find a method fulfilling such 
requirements.
 
In Section \ref{cases} we describe the datasets used for our investigation and  
in Section \ref{filter} we give an overview of state-of-the-art filtering
packages and algorithms, with a focus on radio astronomy.
We also describe the filtering techniques chosen for the analysis 
performed in this paper.
In Section \ref{results} we report an analysis of the best
parameters for each of the filtering methods.
In Sections \ref{noise} and \ref{performance} we test the quality 
and the performance of the filtering algorithms implemented.
In Section \ref{conclusion} we discuss the overall results and conclude that the 
adaptive method is the best solution for our problem.

\section{Test Cases}\label{cases}

In this section, we briefly describe the variety of models and observational datasets
used as test cases. Our sample selection was based on two criteria:
\begin{enumerate}[a)]
\item data cubes with low signal-to-noise features 
      such as tails, extra-planar gas and filaments;
\item clean data cubes, i.e.\ with negligible, or at most minor, artifacts due to
      calibration and imaging effects.
\end{enumerate}

The consequence of the second criterion is that the filtering results presented in the 
next sections will be representative for data cubes mainly affected by 
Gaussian noise. 

\subsection{Models}\label{modeldescription}

\begin{figure}[H]
\centering
\includegraphics[width=0.48\textwidth]{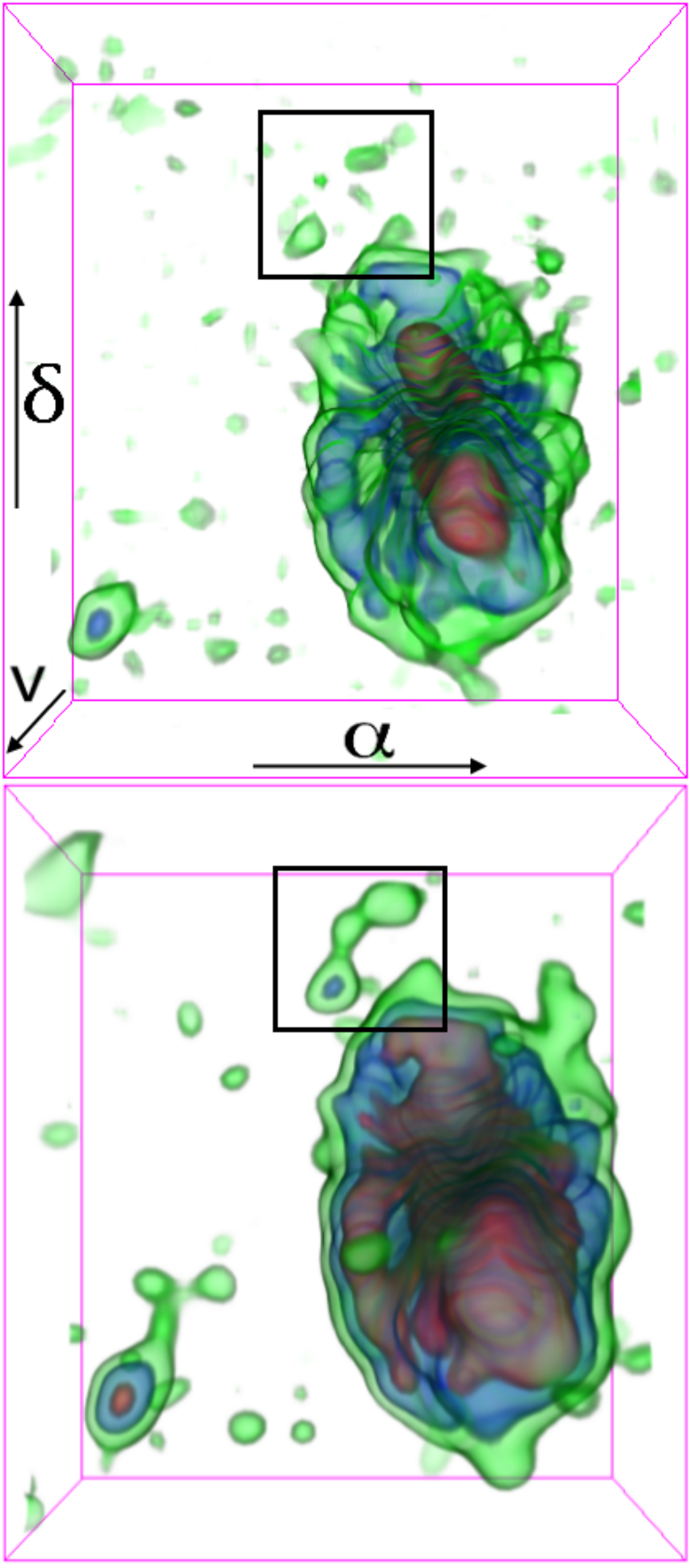}
\caption{views of modelB. 
The upper panel shows a volume rendering of the model
(information regarding the model 
is given in Section \ref{modeldescription}) 
with added Gaussian noise.
The bottom panel shows a volume rendering of the smoothed version 
using an intensity-driven gradient filter with 
parameters $K = 1.0$, $\tau = 0.0325$, $n = 20$ and $C_{x,y,z} = 5$. 
The different colors highlight different intensity levels in the 
data: green, blue and red correspond to 3, 7 and 15 times the $rms$
noise respectively. 
The region of interest, ROI (i.e.\ 
the black box), highlights the 
faint signal, i.e.\ part of a very faint tail.}
\label{ModelBScreeshot}
\end{figure}

We generated several models by taking an existing observation and isolating the detected signal
manually. The object (NGC3359), and hence the model, consists of the H\,{\small I} 
content in 
a spiral galaxy and a small companion, with an incomplete tidal tail-like structure 
between them.  
Gaussian noise has been added with the $\tt{GIPSY}$ \citep{Gipsy, Gipsy1} routine 
RANDOM to produce models with different peak signal-to-noise ratio: 22, 32 and 62,
named ModelA, ModelB 
(see Fig.~\ref{ModelBScreeshot}) and ModelC respectively.
The signal-to-noise properties of ModelB are the closest to the 
observational data shown in this paper. Therefore, ModelB will be used as the main 
reference model.

The data cube size is $6.7 \times 10^5$ voxels. 
The beam size is $\sim 88^{\prime\prime} \times 70^{\prime\prime}$.
The pixel spacing is:
\begin{enumerate}
\item $20^{\prime\prime}$ in Right Ascension (RA), i.e.\ the data up to $\sim 4$ neighboring 
      pixels are correlated.
\item $20^{\prime\prime}$ in Declination (Dec), i.e.\ the data up to $\sim 3$ neighboring 
      pixels are correlated.
\item 8.25 km/s in velocity. The pixels in the velocity direction are not correlated.
\end{enumerate}
These numbers contribute to determining both the optimal width of the filter kernel 
(see Section \ref{results}) 
and the number of independent voxels, that is $N = 5.6 \times 10^4$.

\subsection{NGC4111}\label{NGC4111description}
NGC4111 is one of the brightest lenticular galaxies in the Ursa
Major cluster. The main characteristic of the H\,{\small I} 
emission of NGC4111 is an extended faint filament between the three
sources of the datacube.
The orientation and kinematics of this filament suggest that
the galaxies were tidally stripped from the outer disks by their
nearby companions \citep{VerheijenNGC4111}.

\begin{figure}[H]
\centering
\includegraphics[width=0.48\textwidth]{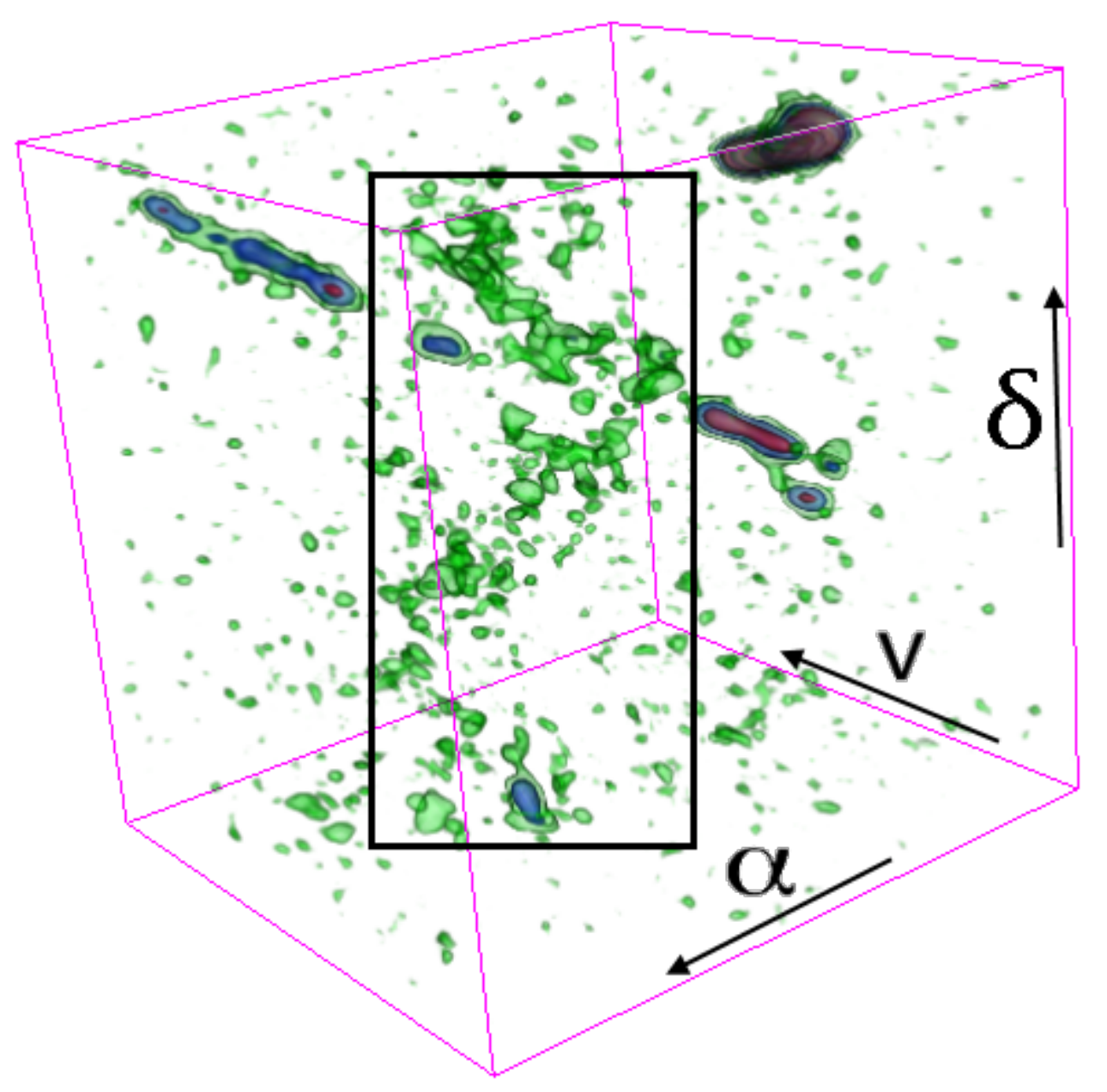}
\caption{A view of the H\,{\scriptsize I} in and around NGC4111 
\citep[information regarding the dataset 
is given in Section \ref{NGC4111description};][in prep]{Eva}. 
The different colors highlight different intensity levels in the 
data: green, blue and red correspond to 3, 7 and 15 times the $rms$
noise respectively.
The region of interest, ROI (i.e.\ 
the black box), highlights the 
faint signal, i.e.\ a faint filament between three galaxies.}
\label{NGC4111Screeshot}
\end{figure}

In Fig.~\ref{NGC4111Screeshot}, we show a volume rendering
of the H\,{\small I} data \cite[][in prep.]{Eva} observed with the
Very Large Array, VLA, telescope. The size of the data cube is
$1.6 \times 10^6$ voxels. 
The beam size is $\sim 45^{\prime\prime} \times 45^{\prime\prime}$.
The pixel spacing is:
\begin{enumerate}
\item $15^{\prime\prime}$ in RA, i.e.\ the data are correlated up to $\sim 3$ neighboring 
      pixels.
\item $15^{\prime\prime}$ in Dec, i.e.\ the data are correlated up to $\sim 3$ neighboring 
      pixels.
\item 5 km/s in velocity. The data are correlated over 2 neighboring 
       pixels because of the use of Hanning smoothing in velocity.
\end{enumerate}
The resulting number of independent voxels is $N = 8.9 \times 10^4$.

\subsection{NGC3379}\label{NGC3379description}
NGC3379 is an elliptical galaxy in the Leo group. The H\,{\small I} 
associated with this galaxy is characterized by a very large, extended tail. 
Part of this tail, such as the wing-shape structure
close to the galaxy, is very faint. 

\begin{figure}
\centering
\includegraphics[width=0.485\textwidth]{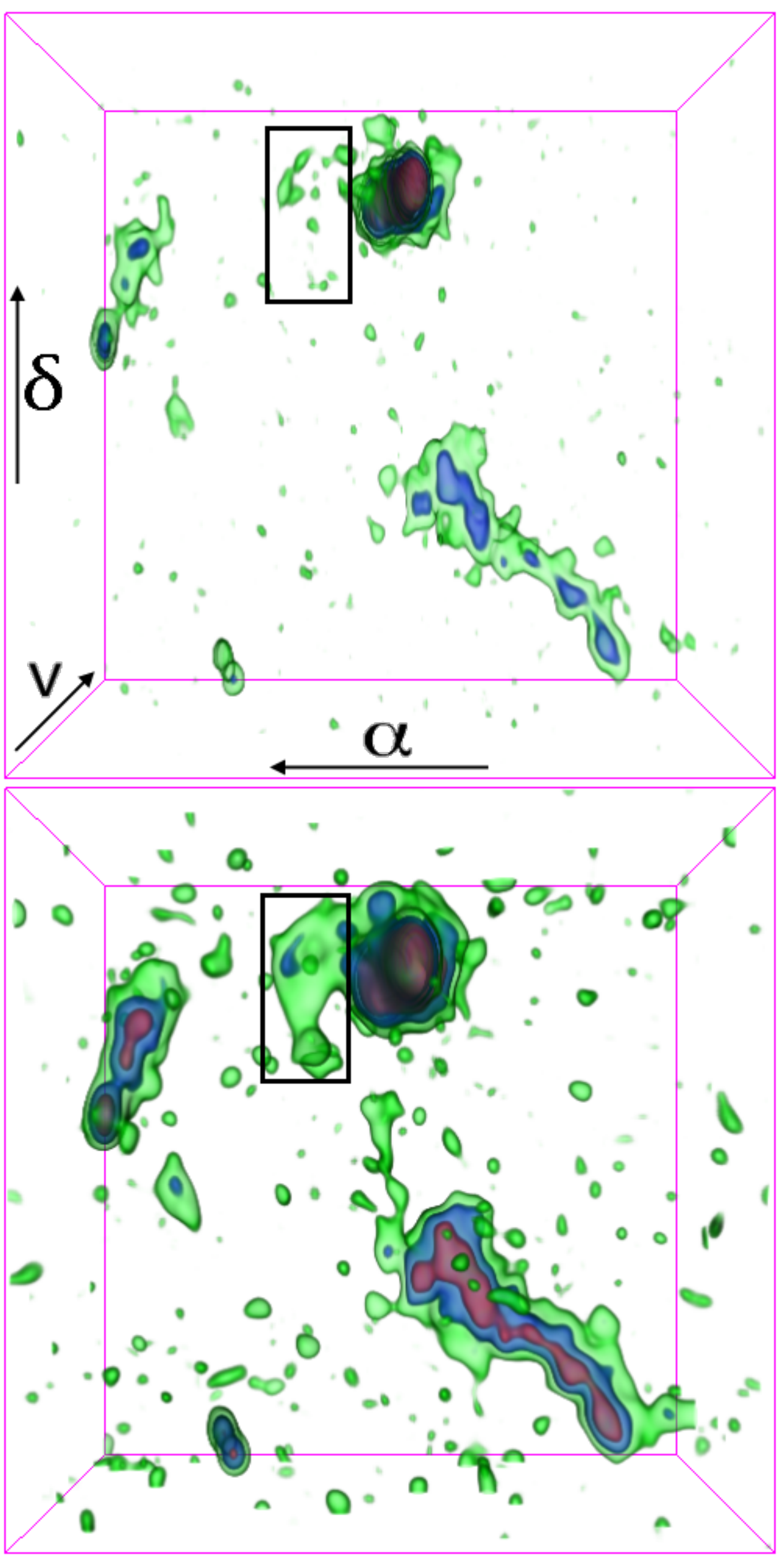}
\caption{Two views of the H\,{\scriptsize I} in and around NGC3379 
\citep[information regarding the dataset 
is given in Section \ref{NGC3379description};][]{atlas3d}.
The upper panel is the volume rendering 
of the original resolution data. 
The bottom panel
shows a volume rendering of the smoothed version 
using an intensity-driven gradient filter with 
parameters $K = 1.5$, $\tau = 0.0325$, $n = 20$ and
$C_{x,y,z} = 5$. 
The different colors highlight different intensity levels in the 
data: green, blue and red correspond to 3, 7 and 15 times the $rms$
noise respectively.
The region of interest, ROI (i.e.\ 
the black box), highlights the 
faint signal, i.e.\ a faint wing-shape tidal structure.}
\label{NGC3379Screeshot}
\end{figure}

In Fig.~\ref{NGC3379Screeshot} we show a volume rendering
of the H\,{\small I} data observed with the
WSRT telescope by \cite{atlas3d}. The size of the data cube is
$3.6 \times 10^7$ voxels. 
The beam size is $81^{\prime\prime} \times 32^{\prime\prime}$.
The pixel spacing is:
\begin{enumerate}
\item $10^{\prime\prime}$ in RA, i.e.\ the data are correlated up to $\sim 8$ neighboring  
      pixels.
\item $10^{\prime\prime}$ in Dec, i.e.\ the data are correlated up to $\sim 3$ neighboring 
      pixels.
\item 8.25 km/s in velocity. The data are correlated over 2 neighboring 
       pixels because of the use of Hanning smoothing in velocity.
\end{enumerate}
The resulting number of independent voxels is $N = 7.5 \times 10^5$.

\subsection{WEIN069}\label{WEIN069description}
The H\,{\small I} data cube of WEIN069 used in this paper 
is a small sub-cube selected from a large mosaic of 48 WSRT
pointings \citep{Mpati}, directed towards a region in the sky
where a filament of the Perseus-Pisces Supercluster (PPScl) crosses the
plane of the Milky Way. The optical counterpart, WEIN069, has been 
observed by \cite{wein}. 

\begin{figure*}
\centering
\includegraphics[width=0.99\textwidth]{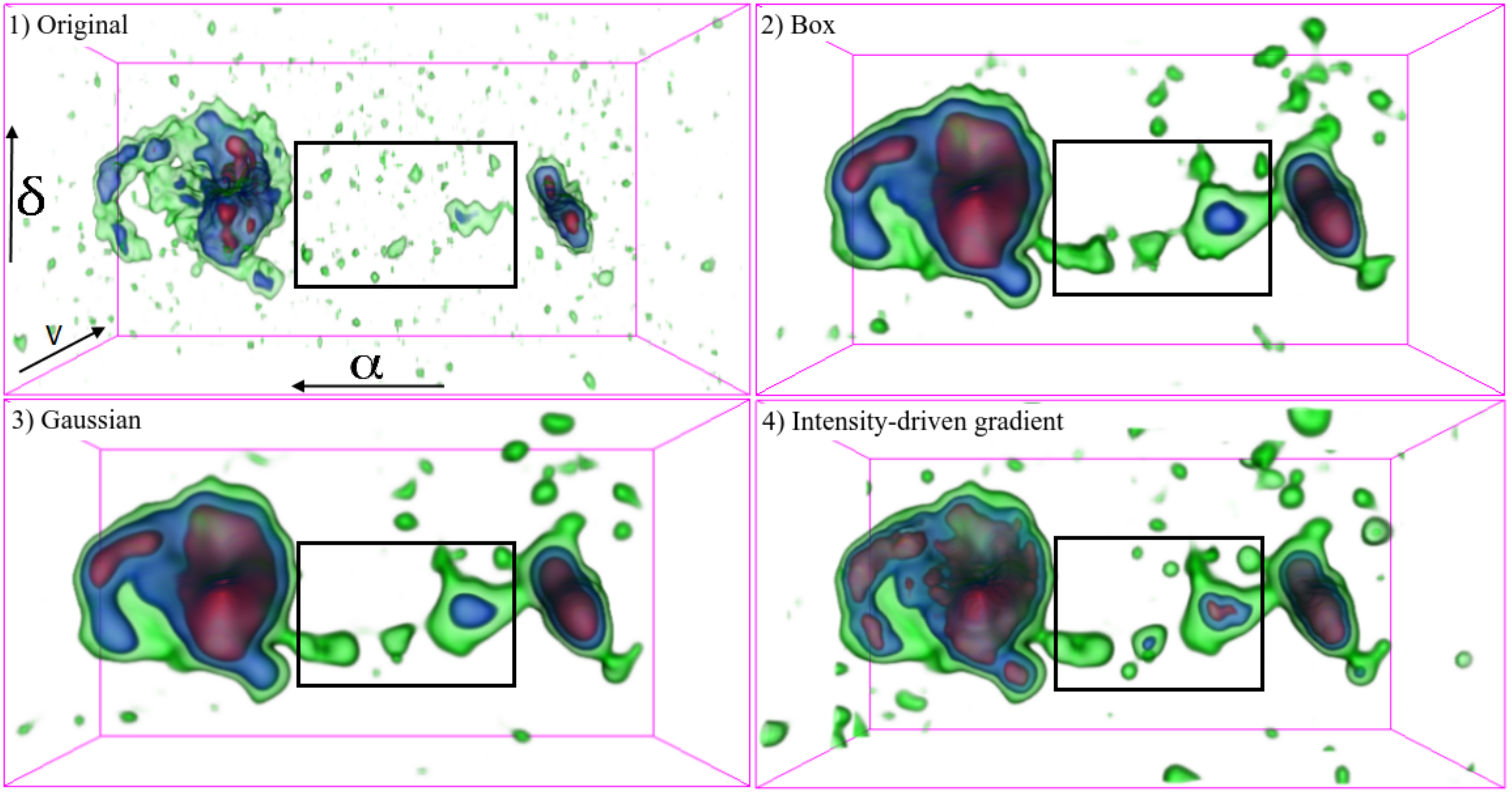}
\caption{The H\,{\scriptsize I} in and around WEIN069 
\citep[information regarding the dataset 
is given in Section \ref{WEIN069description};][]{wein, Mpati}.
The four panels show: 1) a volume rendering 
of the original resolution data; 2) the data filtered with a box filter with 
parameters $N_{x,y,z} = 7$ pixels; 3) the data filtered with a Gaussian
filter with parameters $FWHM_{x,y,z} = 5$ pixels; 4) the data filtered with an
intensity-driven gradient with 
parameters $K = 1.5$, $\tau = 0.0325$, $n = 20$ and
$C_{x,y,z} = 5$.
The different colors highlight different intensity levels in the 
data: green, blue and red correspond to 3, 7 and 15 times the $rms$
noise respectively.
The region of interest, ROI (i.e.\ 
the black box), highlights the 
faint signal, i.e.\ a faint filament between the two companions.}
\label{WEIN069Screeshot}
\end{figure*}

The data cube is shown in Fig.~\ref{WEIN069Screeshot}.
It contains two sources, WEIN069 and a companion,
a tidal tail and a very faint filament that connects the 
two galaxies. Its size is $7.8 \times 10^5$ voxels. 
The beam size is $\sim 15^{\prime\prime} \times 15^{\prime\prime}$. 
The pixel spacing is:
\begin{enumerate}
\item $6^{\prime\prime}$ in RA, i.e.\ the data are correlated up to $\sim 3$ neighboring 
      pixels.
\item $6^{\prime\prime}$ in Dec, i.e.\ the data are correlated up to $\sim 3$ neighboring 
      pixels.
\item 8.25 km/s in velocity. The data are correlated over 2 neighboring 
       pixels because of the use of Hanning smoothing in velocity.
\end{enumerate}
The resulting number of independent voxels is $N = 4.3 \times 10^4$.

\section{Filtering techniques}\label{filter}
In this paper, we focus on interactive
filtering of radio data coupled to interactive visualization.  
The aim is to enhance the manual data inspection, 
in particular of low signal-to-noise H\,{\small I} 
structures.

In the next subsections, we list the filtering algorithms used 
in our analysis in Section \ref{results}. The techniques described 
are aimed to suppress the Gaussian white noise. 
Moreover, such filters perform well for data with the following characteristics:
\begin{enumerate}[i)]
\item signal extended over many pixels;
\item rather small spatial intensity derivatives, i.e.\ no sharp edges. 
\end{enumerate}
Data of H\,{\small I} in and around galaxies fall into this class. A good example is 
presented in Fig.~\ref{WEIN069Screeshot}, one of the data cubes of our sample. 

Artifacts generated by effects such as Radio Frequency 
Interference (RFI), 
errors in the bandpass calibration or in the continuum subtraction
have different statistical properties. Other filter techniques are required to
efficiently characterize these artifacts, tailored to their special spatial and 
spectral signature. 
In this paper we focus on `clean'
data cubes that are considered free from such artifacts.

For a full review of image processing techniques we refer to 
\cite{goyal2012comprehensive, buades2005review, gonzalez2002digital, weeks1996fundamentals}.

It is also worthwhile to mention the following
automated segmentation methodologies (i.e.\ automated 
source mask generation):

\begin{enumerate}
\item $\tt{SoFiA}$ \citep{sofia}: this pipeline has
several tasks for smoothing, source finding and mask optimization. 
A graphical user interface is also available. Three source-finder algorithms
are available: i) a threshold finder; ii) a Smooth and Clip (S-C) finder,
which applies thresholding after smoothing the data with a set of user-specified 
Gaussian kernels and then merges the results; iii) the CNHI finder, 
which performs a threshold rejection Kuiper test on extracted 1-D spectra. 
The completeness and reliability of detected sources 
are evaluated through statistical evaluation of parameters
such as the peak flux, 
total flux, and number of voxels of both positive and negative 
detections. \citep{serra}.
\item $\tt{Duchamp}$ \citep{Whiting}: this pipeline mainly uses a multi-resolution
wavelet transform (specifically the \textit{\`a trous} algorithm; \cite{Starck1994})
for thresholding the data in the wavelet domain. False 
detections are rejected using the false discovery rate technique \citep{Hopkins2002}.
\item $\tt{MAX-TREE}$ \citep{carlinet2012fair}: this is a tree representation 
of the data of which the different nodes are classified 
based on their attributes. These attributes are used to determine the properties
of the node (for more information see 
\cite{teeninga2015improved}). This algorithm has been applied 
both to interactive visualization  \citep{westenberg2007volumetric} and 
optical 2-D data \citep[\textit{MT objects};][]
{teeninga2015improved, teeninga2015improving}. Preliminary experiments 
are also ongoing for H\,{\small I} data 
\citep[\textit{MT source finder};][]{moschini2014, Arnoldus2015}.
\end{enumerate}

\subsection{Box filter}

The mean filter (the box filter) simply consists of replacing
each pixel value in an 
image with the mean value of its neighbors, including itself.
This has the effect of eliminating pixel values 
that are unrepresentative of their surroundings. 

The box filter is a convolution filter.
Like other convolutions, it is based on a 
kernel that represents the shape and size of the neighborhood
to be sampled when calculating the mean. 
Box filtering is most commonly used as a simple method for
reducing noise in an image (see Fig.~\ref{WEIN069Screeshot}). 
However, it has the following drawbacks:

\begin{enumerate}[a)]
\item a single pixel with a strong artifact, such as RFI, can significantly
      affect the mean value of all the pixels in its neighborhood;
\item when the filter neighborhood straddles an edge, the filter will
      blur that edge, leading to a loss of information if the edge is sharp.
      For H\,{\small I} data this is rarely the case:
      the effect is visible around the green
      edges (3 $rms$) of the H\,{\small I} filament (see second panel in 
      Fig. \ref{WEIN069Screeshot}). It is a second order effect which only partially degrades
       the smoothing quality (i.e., the main structure is still visible).  
\end{enumerate}

In general, the box filter acts as a low pass filter and, therefore, 
reduces the spatial intensity derivatives present in the image.
The computational complexity of the box filter is $O(N^3)$, where $N$ is the
number of voxels.

\subsection{Gaussian filter}

The Gaussian filter is a 3-D convolution operator
that is used to denoise images by smoothing.
The kernel is the following Gaussian function:

\begin{equation} 
\mathlarger{G(x,y,z) = A \exp^{-\left(\frac{(x-x_0)^2}{2 \sigma^2_x} 
+ \frac{(y-y_0)^2}{2 \sigma^2_y}
+ \frac{(z-z_0)^2}{2 \sigma^2_z}\right)}},
\end{equation}

\noindent{where the parameters $\sigma_x$, $\sigma_y$, $\sigma_z$
are related to the full width
at half maximum (FWHM) of the peak according to}

\begin{equation} 
FWHM_i = 2 \sqrt{2\ln(2)} \sigma_i, \; i = x,y,z,
\end{equation}

\noindent{which determines the degree of smoothing. 
The 3-D kernel can be also rotated:}

\begin{equation} 
K(x,y,z) = R_z(\theta_z) \; R_y(\theta_y) \; R_x(\theta_x) \; G(x,y,z),
\end{equation}

\noindent{where $R_x$, $R_y$ and $R_z$ are the Euler rotation matrices
corresponding to the three Euler angles 
$\theta_x$, $\theta_y$ and $\theta_z$.}

Once a suitable kernel has been calculated, then the Gaussian
smoothing can be performed using standard 
convolution methods. The computational complexity of the 
Gaussian filter is $O(N^3)$.

When the convolution kernel is isotropic ($\sigma_x = \sigma_y = \sigma_z$),
the convolution can be performed much faster since the equation for the
3-D isotropic Gaussian is separable into
the three axial components. Thus, the 3-D convolution can be
performed with three separate 1-D Gaussian convolutions. 
The computational complexity is then lowered to $O(N)$.

The Gaussian filter outputs a weighted average of each pixel's
neighborhood, with the average weighted more towards 
the value of the central pixels. This is in contrast to
the box filter's uniformly weighted average. Because 
of this, a Gaussian provides gentler smoothing and
preserves edges better than a similarly sized mean filter 
\citep{buades2005review}.
For  H\,{\small I} data, this effect is minor, however
it is possible to observe some small 
differences at the 3 $rms$ level in Fig.~\ref{WEIN069Screeshot}
(in the 3-D views the faint signal and noise at 
the 3 $rms$ level are highlighted in green). These discrepancies 
increase with larger kernels. 

In order to increase the local signal-to-noise ratio
of the very faint signal, both the box and the Gaussian filter 
have to use large kernels for the convolution \citep{buades2005review}. 
The main drawback of these filters is the loss of the spatial 
information with high signal-to-noise ratio, i.e.\ the inner region 
of the galaxy as shown in the second and third panels
in Fig.~\ref{WEIN069Screeshot}. In the next subsection, we will introduce the 
intensity-driven gradient filter which is designed
to deal with this issue by adaptive smoothing depending on the local 
signal-to-noise ratio and structure in the data.

\subsection{Intensity-Driven Gradient filter}\label{gradientFilter}

The gradient filter \citep{perona1990scale} operates
on the differences between neighboring pixels, 
rather than on the pixel values directly. 
The algorithm, known also as \textit{anisotropic diffusion},
uses a diffusion process described by the following
differential equation:

\begin{equation}\label{perona}
\begin{split}
& \frac{\partial I(x,y,z,t)}{\partial t} = S(x,y,z,t) \Laplace I(x,y,z,t) \; + \\
& \nabla S(x,y,z,t) \cdot \nabla I(x,y,z,t),
\end{split}
\end{equation}

\noindent{where $I$ is the intensity of the pixel and $S$ is the diffusion coefficient.
The algorithm was designed for edge detection by choosing:}

\begin{equation}
S(x,y,z,t) = \frac{\mathlarger{1}}{1 +  
\mathlarger{\frac{\left| \nabla I(x,y,z,t) \right|^2}{K^2}}}.
\end{equation}

Instead of having the degree of blurring be 
dependent on the magnitude of the gradient, it can
also be made dependent on other properties,
such as the squared image intensity \citep{perona1990scale, Arnoldus2015}:

\begin{equation}\label{conductivepara}
S(x,y,z,t) = \frac{\mathlarger{1}}{1 +  
\mathlarger{\frac{I^2(x,y,z,t) }{K^2 \; rms^2}}}.
\end{equation}

Substituting equation \ref{conductivepara} in equation \ref{perona}, 
we obtain a diffusion algorithm which preserves 
the edges less well, but it adaptively smooths the pixel intensity
(i.e. more smoothing for lower signal-to-noise ratio). 
The second term of equation \ref{perona} can be neglected as shown by
\cite{perona1990scale} and we use their approach for the discretization of equation \ref{perona}.
The discretized form of this approximation  
for the $i$-th and $i+1$-th iteration is:

\begin{equation}
I_{i+1} = I_i + 
\mathlarger{\tau} \; \frac{C_x \nabla_x I_i + C_y \nabla_y I_i + C_z \nabla_z I_i} {1 + \mathlarger{\frac{I^2_i}{K^2 \; rms^2}}} ,
\end{equation}

\noindent{where the algorithm evaluates this expression $n$ times from $i=0$ to $i=n$.
$I_i = I_i(x,y,z)$,
$\nabla_x I$ indicates the nearest-neighbor differences defined as 
\newline
$\left[I(x+1,y,z) - I(x,y,z)\right] + \left[I(x-1,y,z) - I(x,y,z)\right]$,  $rms$ is the noise level in 
the data cube and $\tau$, $C_x$, $C_y$, $C_z$ and $K$
are input parameters. The input parameters have the 
following upper and lower limits: i) $\tau$ ranges in 
$[0.0025; 0.0625]$; ii) $C_x$, $C_y$ and $C_z$ range in 
$[0; 10]$; iii) $K$ ranges in $[0.5, 10]$. We define the following
default parameters: $K = 1.5$, $\tau = 0.0325$, $n = 20$, $C_x = C_y = C_z = 5$.}

The intensity-driven gradient filter is intrinsically adaptive and is
therefore a very powerful tool for 
investigating low signal-to-noise, extended emission such as tails, filaments
and extra-planar gas. The fourth panel in Fig.~\ref{WEIN069Screeshot}
shows an example of gradient smoothing. In the inner part of
the galaxy (shown in red at levels above 15 $rms$) the full resolution is conserved
remarkably well, while the fainter structure in the
outer part shown in green (i.e. the filament at 3 $rms$) has been enhanced at the
expense of resolution.
A disadvantage of the adaptive smoothing process is that it does not conserve 
the flux scale. The consequence is that the results can 
be used for visualization purposes, but not for quantitative analysis. 
Operations such as calculating column densities,
intensity weighted mean velocities, velocity dispersions etc.,
must be performed on the original data cube or properly convolved versions. The computational 
complexity of the intensity-driven gradient filter is $O(n\;N)$.

\subsection{Wavelet filter}\label{wavelet}
Wavelet transformations are used to obtain a multiresolution 
representation for analyzing
the information content of images. An advantage is that
in the wavelet domain it is easier to discriminate 
the signal from the noise of the image. The 
decomposition process, mathematically reversible, 
defines a multiresolution 
representation \citep[for more information, see][]{mallat1999wavelet}. 
In this paper, we restricted ourselves to wavelet 
transformations using the orthogonal Haar wavelet \citep{daubechies1998factoring}
and the biorthogonal Le Gall 5/3 wavelet  \citep[known also as Cohen-Daubechies-Feauveau 5/3,
CDF 5/3, wavelet;][]{gall1988sub}.

To obtain a wavelet representation, 
we used a \textit{wavelet lifting}
algorithm \citep{daubechies1998factoring}. Wavelet 
lifting consists in applying 
low and high pass filters, corresponding to the 
chosen wavelet, at different resolutions.
At each resolution, the low pass filter generates 
an \textit{approximation band}, $c^l$,
and the high pass filter generates the \textit{detail band}, 
$d^l$, both of length $N/2^l$ elements, 
where $l$ is the value of the decomposition level.
The approximation bands represent the coarse features in the data,
while the detailed bands represent the fine features. The fine features are
the differences between the full resolution data and the new coarse version. The 
detailed bands are used to restore the original data from the coarse resolution. 

The wavelet lifting algorithm is performed in 3 steps:
\begin{enumerate}[1)]
\item Splitting: this step splits a signal into two sets 
of coefficients, those with even and
those with odd index, indicated by $even^{l}$ and $odd^{l}$. 
This is called the \textit{lazy wavelet transform}.
\item Prediction: as the even and odd 
coefficients are correlated, we can
predict one from the other: 
\begin{equation}
d^{i,l+1} = odd^{i,l} - P(even^{i,l}),
\end{equation}
\noindent{where $i$, is the index for the $i$-$th$ array element, 
and the predict operator, $P$, 
in the case of the Haar wavelet, is}
\begin{equation}
P(even^{i,l}) = even^{i,l}.
\end{equation}
\item Update: similarly to the prediction step
\begin{equation}
c^{i,l+1} = even^{i,l} + U(odd^{i,l+1}),
\end{equation}
\noindent{where the update operator, $U$, for the Haar wavelet, is}
\begin{equation}
U(odd^{i,l+1}) = \frac{d^{i,l+1}}{2}
\end{equation}
\end{enumerate}

An image is then denoised by applying thresholding to 
the detail bands. Performing 
wavelet lifting does not require 
additional memory. 
In addition, the computational complexity
of wavelet lifting is $O(N)$
which makes the algorithm extremely fast.

Wavelet lifting has been widely used as a tool for 
image denoising in several fields. 
A practical example of an application of image denoising 
with wavelet transforms in the case of
functional magnetic resonance imaging (fMRI) can be found in \cite{wink2004denoising}.

\begin{figure}[!ht]
\centering
\includegraphics[width=0.49\textwidth]{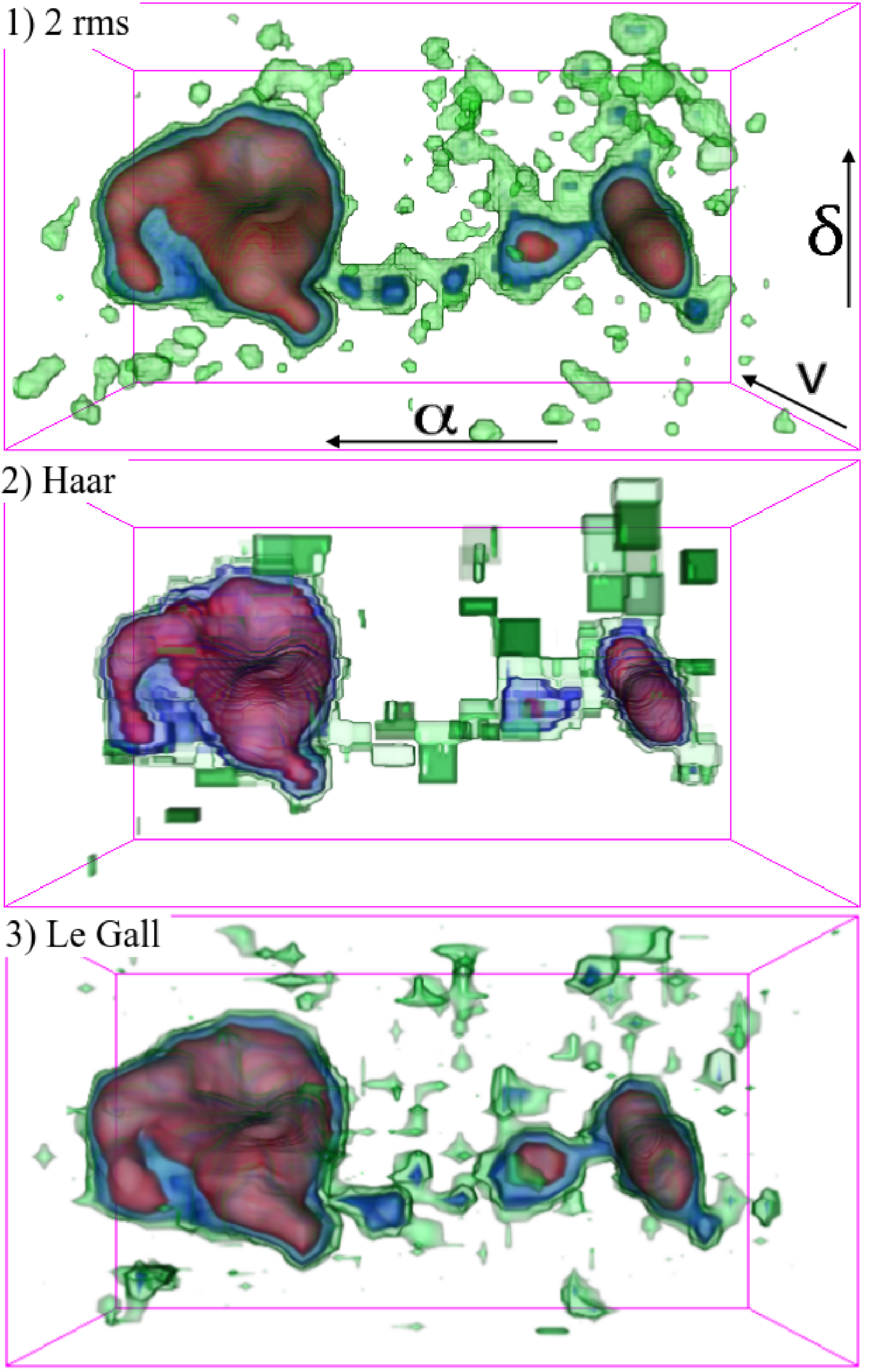}
\caption{Three volume renderings of WEIN069.
In the upper panel, we show the filtered data applying 
a $2 \; rms$ thresholding. In the middle and lower panels
the data are filtered with a Haar and Le Gall wavelet
thresholding filter, respectively. We performed the 
decomposition up to level $l = 3$ for both wavelets, then 
we applied thresholding.
In all the cases, we pre-smoothed the data with the 
same Gaussian filter with parameters $FWHM_{x,y,z} = 5$
pixels. The different colors highlight different intensity levels in the 
data: green, blue and red correspond to 3, 7 and 15 times the $rms$
noise respectively.}
\label{ScreenshotThreshold}
\end{figure}

In Fig.~\ref{ScreenshotThreshold}, we show
filtering results based on the Haar and  
Le Gall 5/3 wavelets. 
We pre-smoothed the data with a Gaussian filter with parameters $FWHM_{x,y,z} = 5$,
then we decomposed the signal up to the third decomposition level and
we finally applied thresholding to the approximation and detail bands. 
We note that, in general, wavelet denoising algorithms 
for suppressing Gaussian noise apply thresholding only to the detail bands.
However, in the case of H\,{\small I} data, we discovered that it is necessary to threshold
both the detail and approximation bands to properly isolate the signal from the noise 
(the signal is extremely faint). 
As a result the algorithm is effectively a thresholding filter.
The values of the thresholding parameters, $t_{l,wavelet}$, used are:
i) $t_{1, Haar} = 0.5$, $t_{2, Haar} = 0.8$ and $t_{3, Haar} = 1.1$
    times the $rms$ of the original data cube for the Haar wavelet;
ii) $t_{1, LeGAll} = 1$, $t_{2, LeGAll} = 1.4$ and $t_{3, LeGAll} = 1.7$
    times the $rms$ of the original data cube for the Le Gall 5/3 wavelet.
Throughout this paper the thresholding parameters will always be defined in units of the 
$rms$ of the original data cube.
Comparing the three panels in 
Fig.~\ref{ScreenshotThreshold}, one can clearly see that the 
wavelet filters remove the noise efficiently with minimal loss of the signal.

The algorithms used have, however, some drawbacks.
The Haar filter looses resolution at low 
signal-to-noise ratio due to the averaging of neighborhood pixels. 
The Le Gall 5/3 filter applies an additional degree of smoothing
and generates clear artifacts as shown in 
Fig.~\ref{ScreenshotThreshold}.

Although the output images obtained by wavelet denoising algorithms are affected by   
artifacts, wavelet thresholding is very promising when
compared to a simple 2 $rms$ thresholding filter. On the other hand, to use
wavelet thresholding effectively we encountered the following complications:

\begin{enumerate}[1)]
\item finding the right multi-level thresholds in the wavelet space 
      is a rather difficult task, which highly depends on the signal-to-noise
      ratio of the faint signal in the wavelet domain;
\item the choice of the decomposition level, $l$, and 
      of the wavelet highly depends on the spatial and velocity extents
      of the unknown faint signal (e.g.\ higher order wavelets may give different results). 
\end{enumerate}

A full investigation to determine the optimal wavelet,
decomposition level and threshold values for denoising H\,{\small I} data with a wide range
of properties will be extremely useful. \cite{Floer} provided an analysis and application 
of wavelet filters for source finding. They demonstrated that  
separating the wavelet analysis of the spatial dimensions from the velocity dimension
increases the filtering quality.
However, their study focused on non-resolved galaxies.
In the case of well-resolved galaxies the presence of faint and unusual
H\,{\small I} structures adds even more complexity to the problem. 
We will discuss this further in Section \ref{conclusion}.   

\section{Optimal filtering parameters}\label{results}

In Section \ref{filter}, we qualitatively illustrated  the
filtering results of applying box, Gaussian, intensity-driven gradient, 
and wavelet lifting algorithms to the WEIN069 data cube (Fig.~\ref{WEIN069Screeshot}).
In this section, we compare quantitatively the box, Gaussian and intensity-driven 
gradient filtering output, for the full sample defined in Section \ref{cases}.
In order to quantify the smoothing quality, we define a diagnostic parameter:
\begin{equation}\label{F_equa}
F =  \frac{|S_{o,f}|}{|S_{o,o}|},
\end{equation}

\noindent{where}
\begin{equation}
\begin{aligned}
& S_{o,f} = \sum_{(x,y,z)} I_{o}(x,y,z) \; M_{f}(x,y,z) \\
& S_{o,o} = \sum_{(x,y,z)} I_{o}(x,y,z) \; M_{o}(x,y,z),
\end{aligned}\label{Sequa}
\end{equation}

\begin{equation}
M_{i}(x,y,z) = \begin{cases}
1  &\text{if $I_{i}(x,y,z) \geqslant 3\, rms_{i}$,\;} i = o, f\\
0  &\text{if $I_{i}(x,y,z) < 3\, rms_{i}$}.
\end{cases}\label{Mfequa}
\end{equation}

In the previous equations, $rms_{i}$ is the root mean square (i.e.\ noise level), $I_{i}(x,y,z)$ is 
the intensity of the pixel at coordinates $(x,y,z)$, the index $o$ refers to the 
original data cube and $f$ to the filtered one. The coordinates $(x,y,z)$ range in a 
ROI sub-cube of a faint signal as shown (with a black box) in Figs. \ref{ModelBScreeshot},
\ref{NGC4111Screeshot}, \ref{NGC3379Screeshot} and \ref{WEIN069Screeshot}.
Moreover, the values of the sums $S_{o,o}$ and $S_{o,f}$, in equation \ref{Sequa}, are 
always calculated on the pixel intensity values of the original data cube.
Therefore, it represents a measurement independent of the filtering technique used. 

The $F$ parameter can range between $[0, M]$ where $M$ is an unknown upper limit 
(see Section \ref{noise}). 
The parameter has a different meaning depending on its range:
\begin{enumerate}[i)]
\item $F \in [0,1]$: the smoothing has washed out the faint signal. This can easily 
       happen using box or Gaussian kernels that are too large.
\item $F \in [1,M]$: the faint signal has been enhanced and the
       number of voxels in the mask $M_f$ is generally larger than in $M_o$. 
       The $F$-value is correlated with the smoothing quality.
       For high values of $F$, the filtered data cube has more signal
       raised over its $3 \; rms$ noise level. 
\end{enumerate}

The error, $\sigma_F$, is propagated as:
\begin{equation}
\begin{aligned}
& \sigma_F =  \sqrt{\left(\frac{\partial F}{\partial S_{o,f}}\right)^2 \sigma^2_{S_{o,f}} \; + \left(\frac{\partial F}{\partial S_{o,o}}\right)^2 \sigma^2_{S_{o,o}}} = \\
& \sqrt{\frac{\sigma^2_{S_{o,f}}}{S^2_{o,o}} 
 + \frac{S^2_{o,f} * \sigma^2_{S_{o,o}}}{S^4_{o,o}}},
\end{aligned}
\end{equation}

\noindent{where $S_{o,o}$ and $S_{o,f}$ are affected by an error
due to the Gaussian noise background equal to}

\begin{equation}
\sigma_{S_{o,i}} = \sqrt{N_i} \, rms_i.  
\end{equation}

In the last equation $i$ is an index which is either $f$ or $o$, $N_i$ is the number of 
independent voxels in the mask $M_i$ and we assumed the $rms_i$ to be constant in the full
data cube.
\begin{table*}[!ht]
  \centering
  \begin{tabular}{ | c | c | c | c |}
    \hline
    data cube &  Filter & Best input parameters & $F$ \\ \hline \hline
    & 1 &  $N_x = 3$ ; $N_y = 3$ ; $N_z = 1$ & $1.1996 \pm 0.0307$ \\ 
	ModelA &  2 & $FWHM_x = 3$ ; $FWHM_y = 3$ ; $FWHM_z = 1$ & $1.1946 \pm 0.0286$ \\
    &  3 &  Haar wavelet ; $l = 2$ ; $t_1 = 0.7$ ; $t_2 = 1.0$  & $ 3.8251 \pm 0.0778$ \\
	&  4 & $K = 1$ ; $\tau = 0.0325$ ; $n = 20$ ; $C_x = 4$ ; $C_y = 5$ ; $C_z = 4$ & $2.1787 \pm 0.0463$ \\  \hline  
    &  1 & $N_x = 3$ ; $N_y = 5$ ; $N_z = 3$ & $1.2975 \pm 0.0196$ \\ 
	ModelB &  2 & $FWHM_x = 3$ ; $FWHM_y = 5$ ; $FWHM_z = 3$ & $1.5959 \pm 0.0228$ \\ 
    &  3 &  Haar wavelet ; $l = 2$ ; $t_1 = 0.7$ ; $t_2 = 0.8$ & $ 2.8515 \pm 0.0394$ \\
	&  4 & $K = 1$ ; $\tau = 0.0325$ ; $n = 20$ ; $C_x = 5$ ; $C_y = 5$ ; $C_z = 4$ & $2.3280 \pm 0.0339$ \\ \hline 
    &  1 & $N_x = 5$ ; $N_y = 5$ ; $N_z = 1$ & $1.0928 \pm 0.0078$ \\ 
	ModelC &  2 & $FWHM_x = 5$ ; $FWHM_y = 5$ ; $FWHM_z = 1$ & $1.2312 \pm 0.0082$ \\  
    &  3 &  Haar wavelet ; $l = 2$ ; $t_1 = 0.9$ ; $t_2 = 0.6$  & $ 1.1062 \pm 0.0072$ \\
	&  4 & $K = 1$ ; $\tau = 0.0325$ ; $n = 20$ ; $C_x = 6$ ; $C_y = 6$ ; $C_z = 4$ & $1.6407 \pm 0.0109$ \\ \hline 
	&  1 & $N_x = 9$ ; $N_y = 7$ ; $N_z = 7$ & $1.9967 \pm 0.006$8 \\ 
	WEIN069 &  2 & $FWHM_x = 7$ ; $FWHM_y = 5$ ; $FWHM_z = 5$ & $2.2576 \pm 0.0076$ \\
    &  3 & Le Gall wavelet ; $l = 3$ ; $t_1 = 0.6$ ; $t_2 = 1.9$ ; $t_3 = 1.45$ & $2.8999 \pm 0.0096$ \\
	&  4 & $K = 1.5$ ; $\tau = 0.0325$ ; $n = 20$ ; $C_x = 6$ ; $C_y = 5$ ; $C_z = 4$ & $2.3392 \pm 0.0081$ \\ \hline 
    &  1 & $N_x = N_y = N_z = 9$ & $3.0789 \pm 0.0032$ \\ 
	NGC4111 &  2 & $FWHM_x = FWHM_y = FWHM_z = 7$ & $3.3057 \pm 0.0034$ \\ 
    &  3 &  Le Gall wavelet ; $l = 3$ ; $t_1 = 1.1$ ; $t_2 = 0.9$ ; $t_3 = 1.2$ & $ 3.7505 \pm 0.0036$ \\
	&  4 & $K = 2$ ; $\tau = 0.0325$ ; $n = 30$ ; $C_x = 5$ ; $C_y = 6$ ; $C_z = 5$ & $2.9665 \pm 0.0031$ \\ \hline 
    &  1 & $N_x = N_y = N_z = 9$ & $5.6655 \pm 0.0078$ \\ 
	NGC3379 &  2 & $FWHM_x = FWHM_y = FWHM_z = 7$ & $5.9252 \pm 0.0081$ \\ 
    &  3 & Le Gall wavelet ; $l = 3$ ; $t_1 = 0.6$ ; $t_2 = 1.15$ ; $t_3 = 1.2$ & $ 6.3993 \pm 0.0233$ \\
	&  4 & $K = 2$ ; $\tau = 0.0325$ ; $n = 30$ ; $C_x = C_y = C_z = 6$ & $5.2800 \pm 0.0072$ \\  \hline 
  \end{tabular}
  \caption{Best runs are reported. We performed    
  the selection evaluating the $F$-values and confirming it by visual inspection. 
  The filter index entries are respectively: 1) box; 2) Gaussian; 3) wavelet lifting thresholding (with Gaussian 
  pre-smoothing); 4) intensity-driven gradient;; 
  The parameters $N$ and $FWHM$ are defined in pixel units. The parameters $t_{l, wavelet}$ are defined in 
  units of $rms$ noise level of the original data cube. The parameters 
  $l$, $K$, $\tau$, $n$ and $C$ are dimensionless.}\label{resulttable}
\end{table*}

We report the values of the $F$ parameter ($F$-values)
in Table \ref{resulttable}, in which 
the best runs and their parameters are reported for each data cube and filter.
The results shown in the table are 
due to a fine-tuning process of the parameter space based both 
on visual inspection of the data and evaluation of the $F$-values. 
The specific input parameter space for each algorithm is:
\begin{enumerate}[1)]
\item box filter: $N_j = 1, 3, 5$ for the Models; $N_j = 5, 7, 9$ for 
      WEIN069, NGC4111 and NGC3379;
\item Gaussian filter: $FWHM_j = 1, 3, 5$ for the Models;
      $FWHM_j = 3, 5, 7$ for WEIN069, NGC4111 and NGC3-379;
\item wavelet filter: $l = 1, 2, 3$;
      \newline
      $t_{1,Haar} = 0.1, 0.3, 0.5, 0.7, 0.9$,
      \newline
      $t_{2,Haar} = 0.4, 0.6, 0.8, 1.0, 1.2$,
      \newline
      $t_{3,Haar} = 0.7, 0.9, 1.1, 1.3, 1.5$,
      \newline
      $t_{1,LeGall} = 0.6, 0.85, 1.1, 1.35, 1.6$,
      \newline
      $t_{2,LeGall} = 0.9, 1.15, 1.4, 1.65, 1.9$,
      \newline
      $t_{3,LeGall} = 1.2, 1.45, 1.7, 1.95, 2.2$;
      \newline
      we also pre-smoothed the data with a Gaussian filter
      with parameters $FWHM_{x,y,z} = 3$ for the models;
      $FWHM_{x,y,z} = 5$ for WEIN069, NGC3379 and NGC4111;
\item intensity-driven gradient filter: $K = 0.5, 1, 1.5, 2$; 
      $n = 20, 30$; $\tau = 0.0325, 0.0625$; $C_j = 4, 5, 6$;   
\end{enumerate}
\noindent where $j = x, y, z$. 
Note that a detailed tuning parameter search can be
performed iteratively at higher resolutions \citep{BergnerSMAS13}. 
However, the input parameter sample used
is accurate enough for finding optimal $F$-values 
and, therefore, for judging which are the best input parameters. 
This has been checked by performing the analysis also with a higher 
resolution sampling of the input parameters.

Moreover, in our parameter space investigation,
we chose to set the rotation parameters for the Gaussian filter, $\theta_i$, to zero 
to reduce the large input parameters space.
This does not introduce a substantial bias in our investigation
because the dependencies of the results on the rotation parameters are 
negligible. In fact, for our sample only filtering results for WEIN069
show a dependence of the $F$-parameter on the 
Euler rotation angles. In the other cases the faint
signal is mainly oriented along one of the primary axes, 
e.g.\ NGC3379, or it has a more complex morphology such as the 
S-shaped filament in NGC4111 or arc-shaped tail in the models. 
As example, in Table \ref{gaussianRotation}, we report the $F$-values of filtering WEIN069
with a rotated Gaussian kernel. The results show that a particular rotation, run III, 
$\theta_y = 340\degree$, increases the $F$ parameter by a factor of $7.5\%$, 
while in run II, $\theta_z = 
340\degree$, it is smaller by a factor of $9.2\%$. This is expected, in fact, since most of the 
faint signal is aligned along a diagonal axis, corresponding to the x-axis rotated 
by 340\degree with respect to the y-axis. Therefore, in run III the kernel is aligned to the
faint signal, while in run II it is perpendicular to it.

\begin{table}[!ht]
  \centering
  \begin{tabular}{| c | c | c | c | c | c |}
    \hline
    Run & $\theta_x (\degree)$  & $\theta_y (\degree)$ & $\theta_z (\degree)$ & $F$ & $\sigma_F$\\ \hline\hline
    
    I & 	0 & 	0 & 	0 & 	2.0701	 & 	0.0071	  \\ \hline
    II &	0 & 	0 & 	340 & 	1.9776	 & 	0.0068	  \\ \hline
    III &	0 & 	340 & 	0 & 	2.1447	 & 	0.0073	  \\ \hline
 
  \end{tabular}
  \caption{The $F$-values applying to WEIN069 a  
  Gaussian filter with parameters $FWHM_x = 7$ pixels, $FWHM_{y,z} = 3$ pixels, 
  $\theta_x$, $\theta_y$ and $\theta_z$.}\label{WEIN069GaussianFilter2Tab}
  \label{gaussianRotation}
\end{table}

For the wavelet filters, we performed a pre-smoothing step with a Gaussian
filter. This was necessary for increasing the signal-to-noise ratio
and providing the optimal results shown in this paper using
as maximum decomposition level $l = 3$. We experienced that, in the case of H\,{\small I} data, 
performing a Haar or Le Gall wavelet analysis beyond the third 
decomposition level gives rise to many artifacts.

In the next section, we will show detailed tests of the $F$-parameter
to establish that this parameter is a reliable estimator of the quality
of the filtering results. In Section \ref{performance} we will present performance
benchmarks of our parallel implementation of the filtering algorithms,
and show that parallelization is necessary to satisfy the 
interactivity requirement defined in Section \ref{intro}.

\section{Noise consideration}\label{noise}

\begin{figure}[!ht]
\centering
\includegraphics[width=0.40\textwidth]{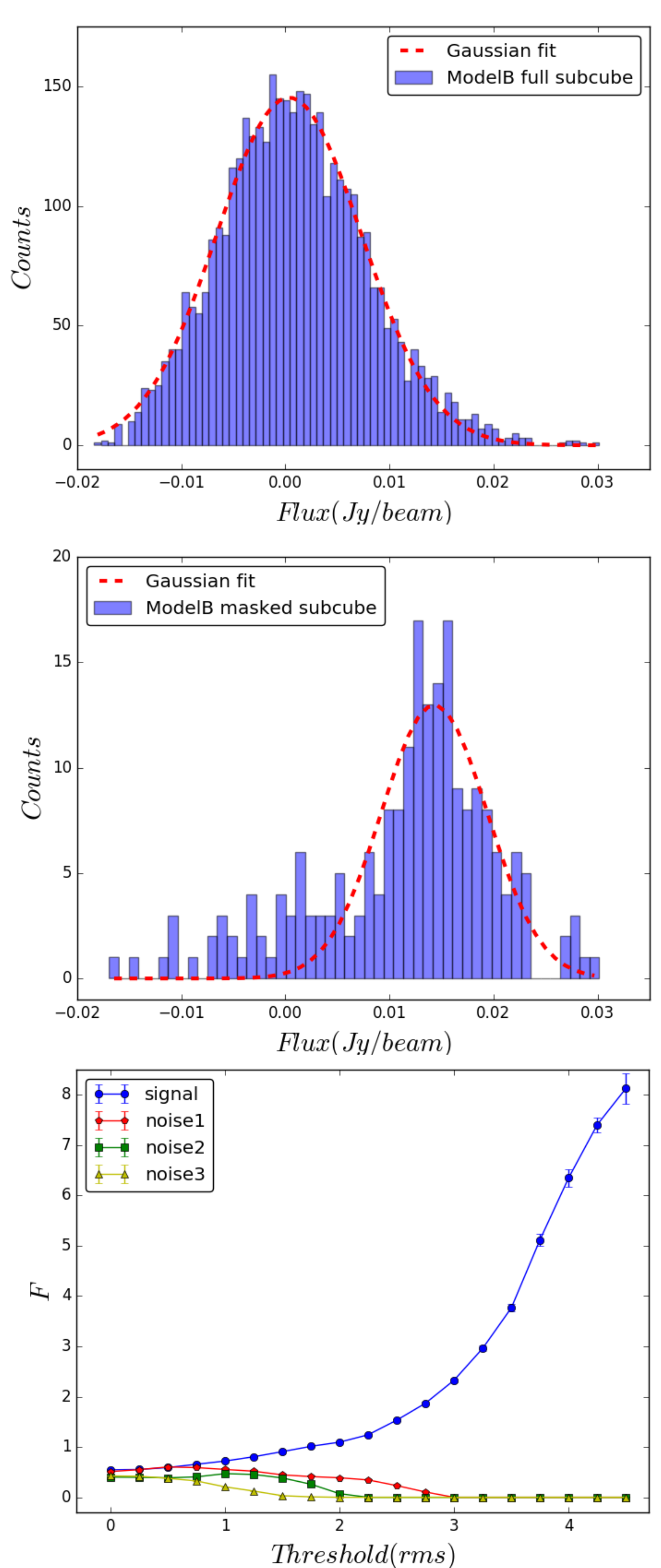}
\caption{The analysis of the histogram of the pixel intensity distribution and 
$F$-parameter for ModelB.
The upper panel shows the histogram of a sub-cube of ModelB.
The sub-cube selection is the ROI, the faint signal, defined in Fig.~\ref{ModelBScreeshot}.
The red curve is a Gaussian fit over the histogram. The fitted parameters are:
$\mu = 3.5\; 10^{-4} \pm 1.0\; 10^{-8}$; $\sigma = 7.0\; 10^{-3} \pm 1.0\; 10^{-8}$;
bins = 75.
In the middle panel, the histogram applying the mask $M_{f}$ 
from run 5 (defined in Table \ref{Noise_table}) on the ROI is shown. 
The output parameters of the fitting are:
$\mu = 1.4\; 10^{-2} \pm 1.6\; 10^{-7}$; $\sigma = 5.1\; 10^{-3} \pm 1.6; 10^{-7}$;
bins = 50.
The lower panel is a plot of the $F$-values calculated from masks obtained by
spanning the thresholding values of the mask $M_f$ from zero to 4.5 \textit{rms}.
The blue line corresponds to the $F$-values calculated on the ROI sub-cube. The red,
green and yellow lines correspond to the $F$-values calculated on three different 
sub-cube, of the same dimension of the ROI, in which there is only noise.}
\label{ModelBHisto}
\end{figure}

In this section, we further investigate the $F$-parameter
defined in equation \ref{F_equa} and its relations 
with the signal and noise.
In fact, the sum over a pure Gaussian noisy sub-cube
is affected by a statistical error 
equal to $\sqrt{N} \;rms$ (i.e.\ the average differs from the zero value).
Moreover, applying the mask calculated in the smoothed data cube, $M_f$, to the original data 
adds further complications: inside the mask there will be a part of the faint
signal (e.g.\ the peak in the histogram of the middle panel in Fig.~\ref{ModelBHisto})
and partially noise (e.g.\ the left wing of the same histogram).

In the lower panel in Fig.~\ref{ModelBHisto} we show a plot of the
$F$-values calculated from masks obtained by
spanning the thresholding value of the mask $M_f$ from zero to 4.5 \textit{rms}.
We performed the calculation both on 
the sub-cube containing the faint signal (ROI 
defined in Fig.~\ref{ModelBScreeshot}) and
three different sub-cubes, of the same dimension as the ROI,
in which there is only noise. In the case of the ROI sub-cube, 
$F$ increases with increased threshold.
Vice versa, for the noise sub-cubes,
$F$ decreases with increasing threshold 
and its value is $\sim 0$ above $2.5 \; rms$.
Note that the threshold used in Section \ref{results}
for the masks $M_i$ is 3 $rms$.

We define also the following parameter:

\begin{equation}
F_{M} =  \frac{|S_{m,f}|}{|S_{m}|},
\end{equation}

\noindent{where}
\begin{equation}
\begin{aligned}
& S_{m,f} = \sum_{(x,y,z)} I_{m}(x,y,z) \; M_{f}(x,y,z) \\
& S_{m} = \sum_{(x,y,z)} I_{m}(x,y,z).
\label{sums_equa}
\end{aligned}
\end{equation}
In these equations, the index $m$ indicates the ModelB
cube without the Gaussian artificial noise.
$S_{m}$ is the integrated flux over the full ROI sub-cube, therefore
$F_{M}$ is the percentage of recovered signal in the mask $M_f$
and it ranges in [0;1]. $M_f$ is defined in equation \ref{Mfequa}.

\begin{figure}[!ht]
\centering
\includegraphics[width=0.48\textwidth]{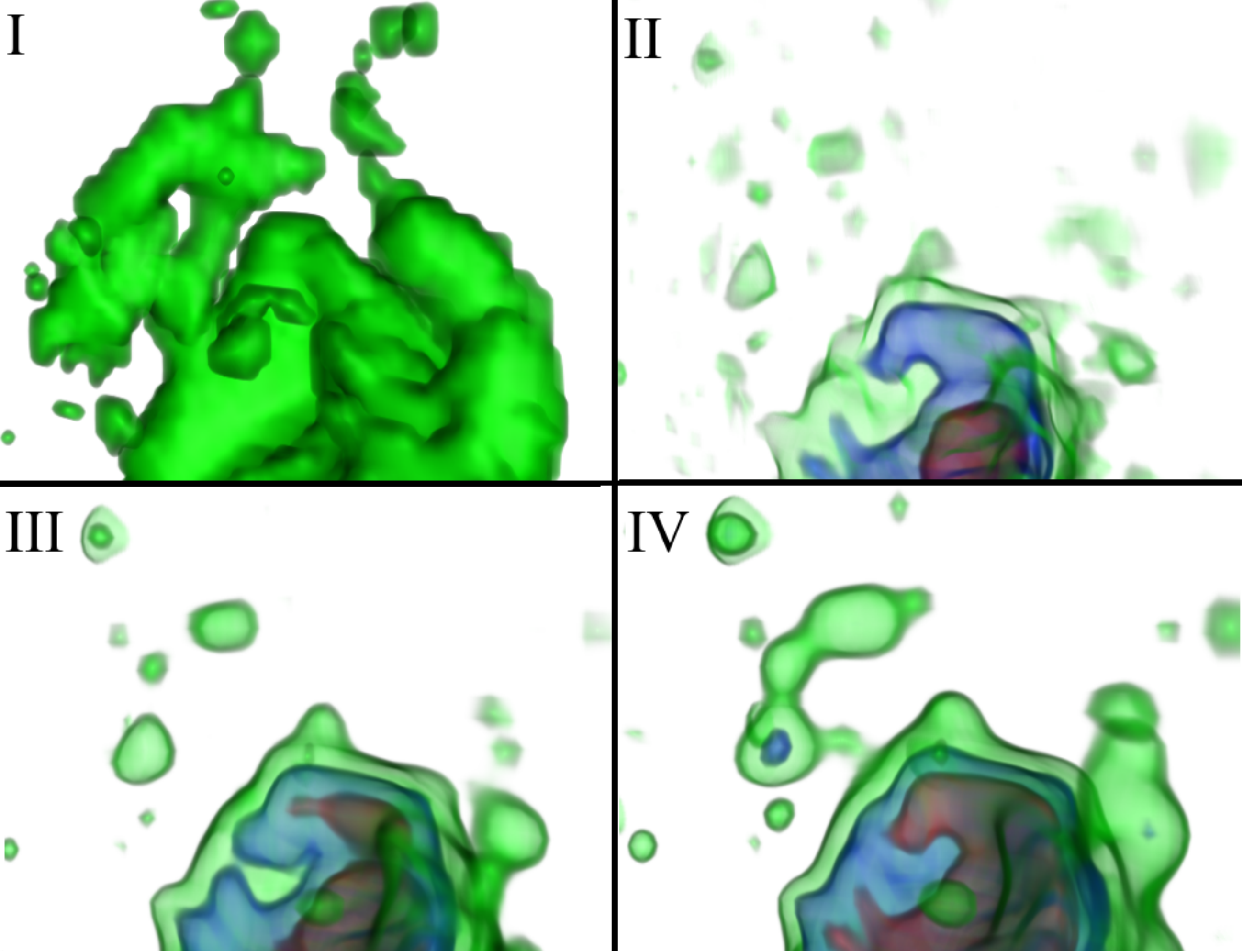}
\caption{In the four views, we look at a zoom of the ROI 
defined in Fig.~\ref{ModelBScreeshot}.
The four panels present the same visualization of four
different data cubes: I) the ModelB without the noise;
II) ModelB; III) the filtering output 
obtained by run 4 ($F = 1.053$; see Table \ref{Noise_table});
IV) the output from run 5 ($F = 2.328$; see Table \ref{Noise_table}).
The different colors highlight different intensity levels in the 
data: green, blue and red correspond to 3, 7 and 15 times the $rms$
noise respectively. The model in the first panel has a 
$rms$ value equal to zero. Therefore, we show only the green level.
}
\label{F_considearion_screenshot}
\end{figure}

In table \ref{Noise_table} we report the $F_{M}$-values obtained by
performing the intensity-driven gradient filter on the ModelB data cube.
The table shows that an increase of the parameter $F$
corresponds to an increase of the parameter $F_{M}$, i.e.\ 
more signal has been recovered in the smoothing process. 
This is also supported by 
visual inspection of the filtered data cubes. In Fig.~\ref{F_considearion_screenshot}, 
the faint signal is clearly enhanced for higher values of $F$ and $F_{M}$.

\begin{table}[!ht]
  \centering
  \begin{tabular}{| c | c | c | c | c | c | c | c |}
    \hline
     Run & $K$ &  $\tau$ & $n$ & $F$ & $F_{M}$ \\ \hline\hline
    
 	1 & 0.5 & 	0.0325 & 	20 & 	$2.128	\pm  	0.032$	& $0.237 \pm  0.004$  \\ 
 	2 & 0.5 & 	0.0625 & 	20 & 	$1.322	 \pm 	0.021$	& $0.149  \pm 0.003$  \\ 
 	3 & 0.5 & 	0.0325 & 	30 & 	$2.193	 \pm 	0.033$	& $0.263 \pm   0.003$  \\ 
 	4 & 0.5 & 	0.0625 & 	30 & 	$1.053	 \pm 	0.018$	 & $0.129 \pm 0.003$ \\ \hline
 	5 & 1 & 	0.0325 & 	20 & 	$2.328	 \pm 	0.034$	 & $0.364  \pm 0.003$  \\ 
 	6 & 1 & 	0.0325 & 	30 & 	$2.148	 \pm 	0.031$	 & $0.334 \pm  0.003$ \\ \hline
 	7 & 1.5 & 	0.0325 & 	20 & 	$1.722	 \pm 	0.025$	& $0.259 \pm  0.002$  \\ 
 	8 & 1.5 & 	0.0325 & 	30 & 	$0.633	 \pm 	0.010$	& $0.071 \pm  0.001$  \\ \hline
 	9 & 2 & 	0.0325 & 	20 & 	$0.701	 \pm 	0.011$	& $0.079 \pm  0.001$  \\ 
 	10 & 2 & 	0.0325 & 	30 & 	$0.263	 \pm 	0.006$	& $0.024 \pm  0.001$   \\ \hline

  \end{tabular}
  \caption{The $F$ and $F_{M}$-values for applying
  an intensity-driven gradient filter with parameters 
  $K$, $\tau$, $n$, $C_{x,y,z} = 5$ to ModelB.}
  \label{Noise_table}
\end{table}

We performed the same analysis  
for ModelA and ModelC, with similar results as
the analysis performed on ModelB.

We conclude that the $F$-values are reliable 
and the noise effects on the $F$-values, 
calculated at the 3 \textit{rms} noise level, are
minor or negligible.

\section{Performance}\label{performance}

In this section, we provide measurements of
the performance of the codes\footnote{The codes are publicly available at 
\url{https://github.com/Punzo/SlicerAstro}} used in this paper.
We performed the benchmark on 
a Linux laptop (Ubuntu 15.10) equipped with:
\begin{enumerate}[-]
\item an Intel i7 2.60 GHz CPU,
\item 16 GB of DDR3 1.6 GHz \textit{random access memory}, RAM,
\item an Intel HD Graphics 4600 \textit{graphics processing unit}, GPU, (it can use up to 1.7 GB of the RAM),
\item an NVIDIA GeForce GTX860M GPU (with 2 GB of  \textit{dynamic random-access memory}, DRAM).

\end{enumerate}

We define the speedup, $S$, as

\begin{equation}
S(N) = T_1(N) / T_p(N),
\label{performanceEqua}
\end{equation}

\noindent{where $T_1$ is the execution time exploiting only one CPU core, 
$T_p$ is execution time of the parallelized code and $N$ the number of voxels.
The codes are parallelized both on CPU (OpenMP) and
GPU (OpenGL). In the case of the GPU implementation, 
the I/O times (i.e.\ times for sending the data to the GPU 
and to getting the results back) are included in the term $T_p$.

\begin{figure*}[!ht]
\centering
\includegraphics[width=1.\textwidth]{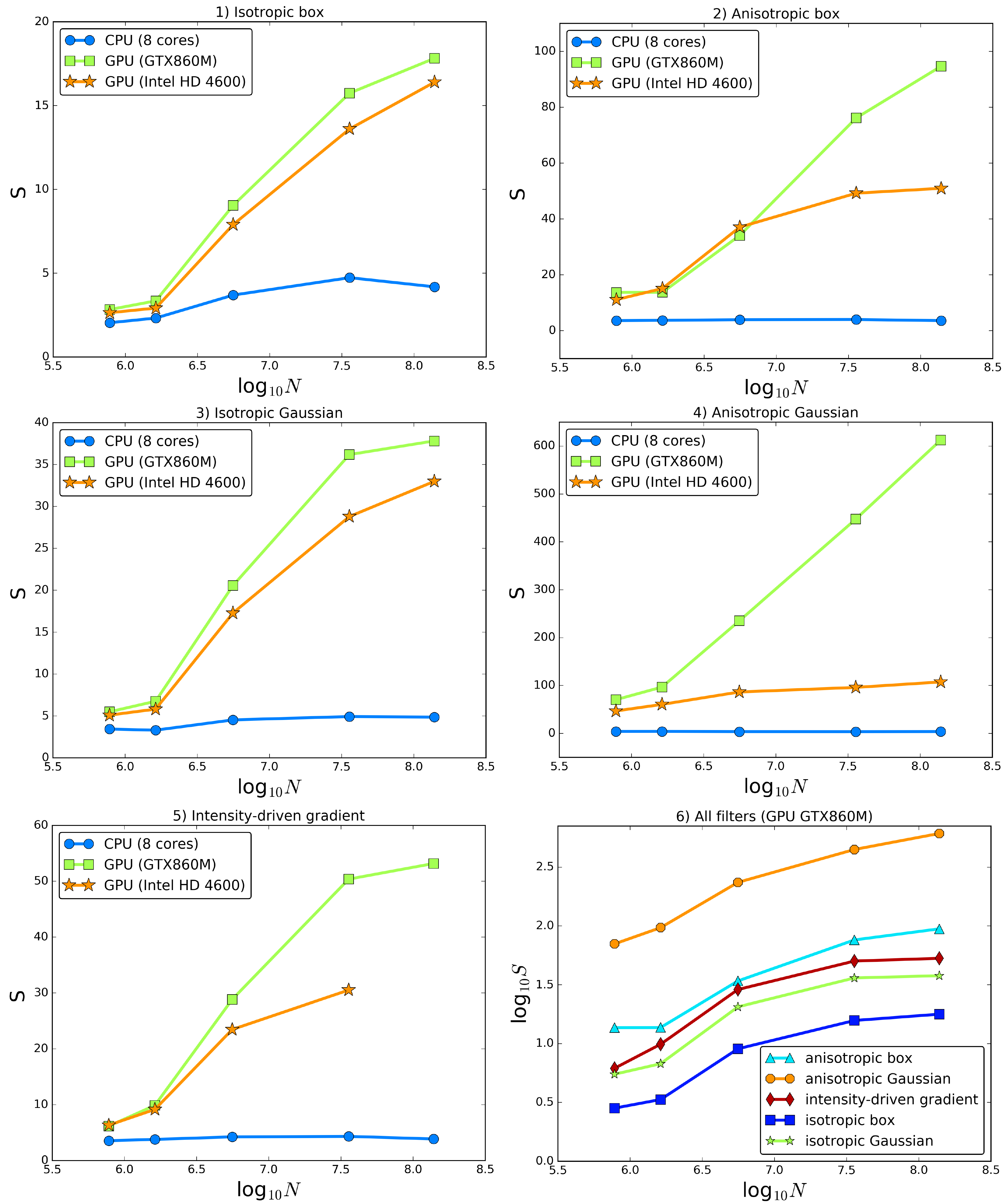}
\caption{The values of speedup of the parallelization of the various filter
algorithms are shown: 
1) upper-left panel, isotropic box; 
2) upper-right panel, anisotropic box;
3) middle-left panel, isotropic Gaussian;
4) middle-right panel, anisotropic Gaussian;
5) bottom-left panel, intensity-driven gradient;  
6) bottom-right panel, comparison of the GPU (GTX860M) implementation 
of all filters. 
The values of the speedup $S$ are calculated using 
equation \ref{performanceEqua}.
$N$ is the number of voxels. The values of the input parameters are defined in section 
\ref{performance}.}
\label{performancefig}
\end{figure*}

We report the speedup results in Fig.~\ref{performancefig}, using the following values
for the input parameters of the filters:
\begin{enumerate}[1)]
\item isotropic box: $N_x = N_y = N_z = 3$ pixels;
\item anisotropic box: $N_x = 3$ pixels and $N_y = N_z = 5$ pixels; 
\item isotropic Gaussian: $FWHM_x = FWHM_y = FWHM_z = 3$ pixels 
      and $\theta_x = \theta_y = \theta_z = 0\degree$
\item anisotropic Gaussian: $FWHM_x = 2$ pixels, $FWHM_y = FWHM_z = 3$ pixels
      and $\theta_x = \theta_y = \theta_z = 0\degree$;
\item intensity-driven gradient: $K = 1.5$, $\tau = 0.0325$, $n = 20$
      and $C_x = C_y = C_z = 5$.
\end{enumerate}
 
A number of conclusions can be drawn from Fig.~\ref{performancefig}.
First, the values of the speedup $S$ for the CPU (8 cores) implementation
      for the various filters at different $N$ are
      $\lesssim 4$. Therefore, the execution time for both the anisotropic box
      and Gaussian filter at large $N$ ($\sim 10^8$) is rather long: 
      $1$ $\rm{min}$ and $10$ $\rm{min}$, respectively. 
      The isotropic Gaussian filter at large $N$
      takes $14.5$ $\rm{s}$ using 8 cores, while the execution time for
      the isotropic box filter is $2$ $\rm{s}$, and $56$ $\rm{s}$ for
      the intensity-driven gradient filter. 
      We also compared our optimized CPU implementation of 
      the isotropic Gaussian filter with the one provided by
      the Insight Segmentation and Registration Toolkit 
      \citep[$\tt{ITK}$;][]{ITK}. Our implementation, using the same number 
      of CPU cores (i.e. 8), showed a speedup by a factor of $3$ over the $\tt{ITK}$ version.}

Secondly, very large values of $S$ are found for the GPU implementation of the 
      anisotropic box and Gaussian filters.
      For example, in the case of 
	  anisotropic Gaussian filtering of NGC2841
	  \citep[$N = 1.4 \times 10^8$, i.e.\ 529 MBytes;][]{Walter},
	  the execution time improves from $35$ $\rm{min}$, 
	  using one CPU core, to $3.5$ $\rm{s}$ exploiting the GTX860M.

Thirdly,  the values of $S$ for the GPU implementation are smaller 
      for the isotropic box and Gaussian filters than for their anisotropic counterparts. 
	  The GPU execution time for the isotropic Gaussian filter with $N = 1.4 \times 10^8$
	  is $1.8$ $\rm{s}$ and therefore a factor of 2 smaller than for the
      anisotropic Gaussian filter.
 
On the other hand the GPU execution time for the intensity-driven gradient filter
      with $N = 1.4 \times 10^8$ is $3.1$ $\rm{s}$, as compared to $4$ $\rm{min}$ with a
      single CPU core.

When examining the behavior in relation to the number of voxels the following conclusions can be drawn. For a data cube with a small number of voxels ($N \sim 10^6$)
      the $S$ values of the GPU implementation for the isotropic box, 
      isotropic Gaussian and intensity-driven 
      gradient filters are close to the 8 CPU cores performances. This is to be expected as 
      for small $N$ it is not possible to fully load the GPU and properly exploit 
      all the cores. However, up to $5 \times 10^6$ voxels, all filters, when using
      the GPU, reach the kind of performance that allows interactive work (maximum execution time, exploiting the GTX860M,
      is $T_p < 0.3 s)$.

Finally, wavelet lifting is a very fast algorithm: 
the maximum execution time (using the Haar wavelet and a value of $l = 3$), exploiting one CPU core,
for filtering a data cube with up to $10^8$ voxels is $T_p < 5.1$ $s$.
Therefore, we did not implement a GPU version. 
Moreover, the implementation of such parallelization is rather 
challenging mainly because of the memory handling on the GPU. A $\tt{CUDA}$ 
implementation was developed by \cite{Wladimir}
giving a speedup of $\sim 10$ with respect to their optimized CPU implementation.
This is a large improvement 
respect to previous works (e.g. \cite{Wong, Tenllado}).

Defining $\eta = 4 \; N$ Bytes as the
RAM usage for a given data cube,
the memory requirements for each of the filter codes are:
\begin{enumerate}[A)]
\item CPU implementations of the box, Gaussian and intensity-driven gradient filters:
      one permanent $\eta$ on the RAM for storing
      the final results and one temporary $\eta$ for storing partial
      run-time results, so a total memory requirement of $2\eta$ RAM;
\item CPU implementation of the wavelet filters: one permanent
      $\eta$ on the RAM for storing the final results;       
\item GPU implementation of the box, Gaussian and intensity-driven gradient 
      filters: one permanent $\eta$ on the
      RAM, one temporary $\eta$ on the RAM and two temporary $\eta$ on the DRAM, so a
      total memory requirement of $2\eta$ RAM and $2\eta$ DRAM;
\end{enumerate}

In summary, a machine with 16 GB of memory can easily accommodate a $\sim 4$ GB dataset 
when using the box, Gaussian or intensity-driven gradient filter (in case of the GPU implementation
at least 8 GB of DRAM are needed).

\begin{table}[!ht]
  \centering
  \begin{tabular}{ | c | c | c |}
    \hline
      Filter & Hardware & $F$ \\ \hline\hline
	 1 & CPU & 1.7534 $\pm$ 0.0061 \\ 
	   & GPU & 1.7217 $\pm$ 0.0060 \\ \hline
	 2 & CPU & 1.8591 $\pm$ 0.0064 \\ 
	   & GPU & 1.8182 $\pm$ 0.0063 \\ \hline 
	 3 & CPU & 1.6953 $\pm$ 0.0059 \\  
	   & GPU & 1.5386 $\pm$ 0.0054 \\ \hline 
	 4 & CPU & 1.8848 $\pm$ 0.0065 \\ 
	  & GPU & 1.7760 $\pm$ 0.0062 \\ \hline 
	 5 & CPU & 2.3416 $\pm$ 0.0083 \\ 
	  & GPU & 2.2704 $\pm$ 0.0079 \\ \hline 

  \end{tabular}
  \caption{The $F$-values relative to both the CPU and
  GPU filtering implementation of the filters applied to WEIN069. The filter index 
  entries are respectively: 
  1) isotropic box;
  2) anisotropic box; 
  3) isotropic Gaussian;
  4) anisotropic Gaussian; 
  5) intensity-driven gradient. The values of the input
  parameters are defined in section 
  \ref{performance}. }\label{FCPUGPU}
\end{table}

For the GPU implementation, we
chose the shader para-digm (OpenGL), over other computational 
scientific SDK (CUDA or OpenCL), for its compatibility 
with all the GPU vendors.  
Moreover, OpenGL is present in any operating system, 
which simplifies the distribution of the software.
The drawback is that the computations
performed with OpenGL have relatively less precision.
For H\,{\small I} data this is not an issue: the 
scalar range of the pixel intensities is relatively small
and float precision is sufficient for the 
calculations required by the algorithms. 
In fact, the differences between the CPU and GPU 
filtered data cubes are unnoticeable: 
in Table \ref{FCPUGPU}, we compare the GPU
methods smoothing quality to the CPU ones
calculating the $F$-values, and the differences between
the two implementations are less than $5\%$.

In the next section, we will summarize 
and discuss the results presented in the previous sections
focusing on their applicability to visualization.

\section{Discussion and conclusions}\label{conclusion}

Future blind surveys of H\,{\small I} will deliver a 
large variety of data in terms both of the number of galaxies and
additional complex features such as tails, extra-planar gas and
filaments. 
These faint structures can be found in nearby medium/high 
resolved galaxies (e.g. Model and WEIN069 data cube)
and groups of non-resolved galaxies  
(e.g. NGC-3379 and NGC4111).
They have a very low signal-to-noise
ratio of $\sim 1$, but are extended over many pixels.
Efficiently separating such signals from the noise
is not straightforward (visual examples are shown in sections
\ref{cases} and \ref{filter}).
Moreover, in the case of APERTIF and ASKAP, 
it is estimated that tens of such sub-cubes will be collected weekly \citep{duffy}.
This is a large volume of data, and a coupling between the filtering algorithms
shown in this paper and 3-D visualization can enhance
the inspection process of large numbers of galaxies and masks 
provided by source finder algorithms.

In Section \ref{filter}, we reviewed state-of-the-art
filtering algorithms. We qualitatively illustrated  the
filtering results using several methods. 
We then performed a visual inspection of the filtering results, followed by a systematic 
quantitative analysis of the algorithms in Section \ref{results}. 

First, we extensively investigated  the 
parameter space of the input parameters (i.e.\ the extension
and shape of the kernels) of the box and Gaussian filters by applying them to several test 
data cubes. In Table \ref{resulttable}, we indicated the
best filtering runs and their input parameters.
As criterion for selecting the best runs we used the $F$-value, our smoothing 
quality control parameter defined in Sections \ref{results} and \ref{noise},
requiring $F$ to be large.
Thereafter, we confirmed the selection by visually inspecting the 
filtered output data cube.
Table \ref{resulttable} highlights, for our sample, that finding
the input parameters of the best runs is not straightforward.  
In fact, the box and Gaussian kernels are highly dependent on the spatial 
and velocity extents, and the signal-to-noise ratio of the unknown faint signal.
Note that the Gaussian smoothing gives better results
than the box smoothing, because a gentler smoothing preserves better
the shape of the data (the differences are clearly visible in the second
and third panels in Fig.~\ref{WEIN069Screeshot}).
Two examples which suffer from these limitations are:
\begin{enumerate}[1)]
\item ModelB: very faint signal (signal-to-noise ratio $\sim$ 1) with limited extent; 
\item NGC~4111: very extended, relatively faint, signal.
\end{enumerate}
In the first case large kernels are necessary to considerably enhance the very low 
level signal. Large kernels (e.g.\ for the box filter $N_i > 5 $ and for the Gaussian filter 
$FWHM_i > 3$) will, however, wash out the signal because it is not coherent at 
such large scales. In the second case,
very large kernels ($N_i = 9$ for the box filters and $FWHM_i = 7$ for the Gaussian filter) 
provide the best smoothing and the maximum $F$-values. Such kernels drastically 
reduce, however, the spatial and velocity resolution of the data. 

The optimal dimensions of the box and Gaussian kernels strongly depend 
on the extent of the signal and the signal-to-noise ratio. 
The quite different, best input parameters of ModelA, ModelB and ModelC, 
with their different signal-to-noise ratios, illustrate this clearly.
For example, the best runs for modelB use larger kernels in the $y$ direction compared to 
the other models. The optimal kernels for smoothing ModelA and ModelC
have, on the other hand, a very narrow $z$ component. This is expected as a higher noise level
hides the signal and modifies the overall shape of the signal itself 
(i.e. the faintest parts will disappear into the noise).

Second, we analyzed wavelet filters in detail. Our
investigation focused on thresholding the data in the wavelet domain.
We performed the filtering operation exploiting a wavelet lifting algorithm.
Two main wavelets have been
used: the Haar and the Le Gall wavelet.
Wavelet lifting is a powerful technique, but unfortunately it generates artifacts 
undesirable for our visualization purposes (see 
Fig.~\ref{ScreenshotThreshold}). The filtering results give very high
values of the $F$-parameter as shown in Table \ref{resulttable}. The wavelet thresholding
filter, however, requires a thorough investigation of the main parameters
(choice of the basic wavelet, maximum number of levels for 
wavelet decomposition, thresholding values for each decomposition level) for obtaining an 
optimal denoising of the data. We consider this a drawback for user-friendly visualization purposes.

The optimal input parameters 
reported in Table \ref{resulttable} vary for each data cube
of our sample. The thresholds parameters, $t_{l, wavelet}$, have strong dependencies  
on the choice of the wavelet and the signal-to-noise ratio of the faint signal. 
Moreover, the choice of the optimal wavelet and decomposition level, $l$, 
depends on the extent of
the faint structure. For example, the arc-shape structure in the ModelB is very thin along the velocity 
direction (few channels). Therefore, a the Haar wavelet and $l = 2$ are the optimal choice, 
while the Le Gall wavelet and a higher decomposition level, $l = 3$, provide the 
optimal filtering results for WEIN069, NGC3379 and NGC4111, because these data shows 
a more extended component.

Filtering with a higher order wavelet than Le Gall may
give optimal results without requiring a pre-smoothing step.
However, we showed that the choice of the wavelet is constrained
by the unknown extent of the faint signal. For example, very high-order wavelets are not 
optimal for filtering the models.

Using different decomposition levels in each spatial
and velocity dimension \citep[or a tree structure, e.g. $\tt{Octree}$;][]{octree}
may also improve the filtering quality. However, 
in the case of morphological complex resolved galaxies
this approach is rather difficult. For example, it is necessary to determine
the optimal levels of decomposition for each dimension and these depend on the
signal extent and signal-to-noise ratio as well.
This is analogous to the issue of finding the optimal kernel for
the box and Gaussian filters.    

Applying wavelet decomposition and thresholding the approximation bands,
as shown in Section \ref{wavelet}, is effectively a segmentation of the data.
Though efficient, the disadvantage is that it also 
eliminates very low signal-to-noise emission if the thresholding 
parameters are not properly tuned to the data.
Since our aim is to couple filtering techniques to visualization, 
thresholding techniques are not favored as they
limit the interactive visual data exploration.

Third, we implemented a modification of the diffusion filter:
the intensity-driven gradient filter (see \ref{gradientFilter}). 
This smoothing algorithm has adaptive characteristics which helps
in preserving the smaller scale structure of the signal, thus avoiding the limitations
of the box and Gaussian filters. 
The parameters of  
intensity-driven gradient filter mainly depend on the 
signal-to-noise ratio of the emission, which we found to be quite 
similar for the objects studied here. In fact, the intensities of the 
majority of the voxels of the faint signal are between 1 and 2 $rms$.
For example, in Section \ref{cases}, we illustrated
3-D visualizations of
the output of the intensity-driven gradient filter
with default parameters ($K = 1.5$, 
$\tau = 0.0325$, $n = 20$ and
$C_{x,y,z} = 5$) for two very different objects (WEIN069 and NGC3379). 
In both cases, the smoothing is successful in bringing out the low 
signal-to-noise structures. 
In fact, in the case of the gradient filter,
the $F$-values of the best runs, 
reported in Table \ref{resulttable}, do not differ
more than $15\%$ from the runs with default parameters.

The main input parameters ($K$, $\tau$ and $n$)
of the best filtering results for the three models in Table \ref{resulttable}
do not vary. The peak signal-to-noise ratio of ModelC is $\sim 3$ times higher 
than that of ModelA. Therefore, the dependencies of the input parameters of the 
intensity-driven gradient respect to the signal-to-noise ratio are not stiff functions.

We conclude that the intensity-driven gradient is the
most promising filter because it preserves the detailed structure of the
signal with high signal-to-noise ratio ($> 3$)
at the highest resolution, while smoothing only the faint part of the signal 
(S/N $< 3$). Moreover, the input parameters need only minimal tuning to the signal itself.

\begin{figure}[!ht]
\centering
\includegraphics[width=0.48\textwidth]{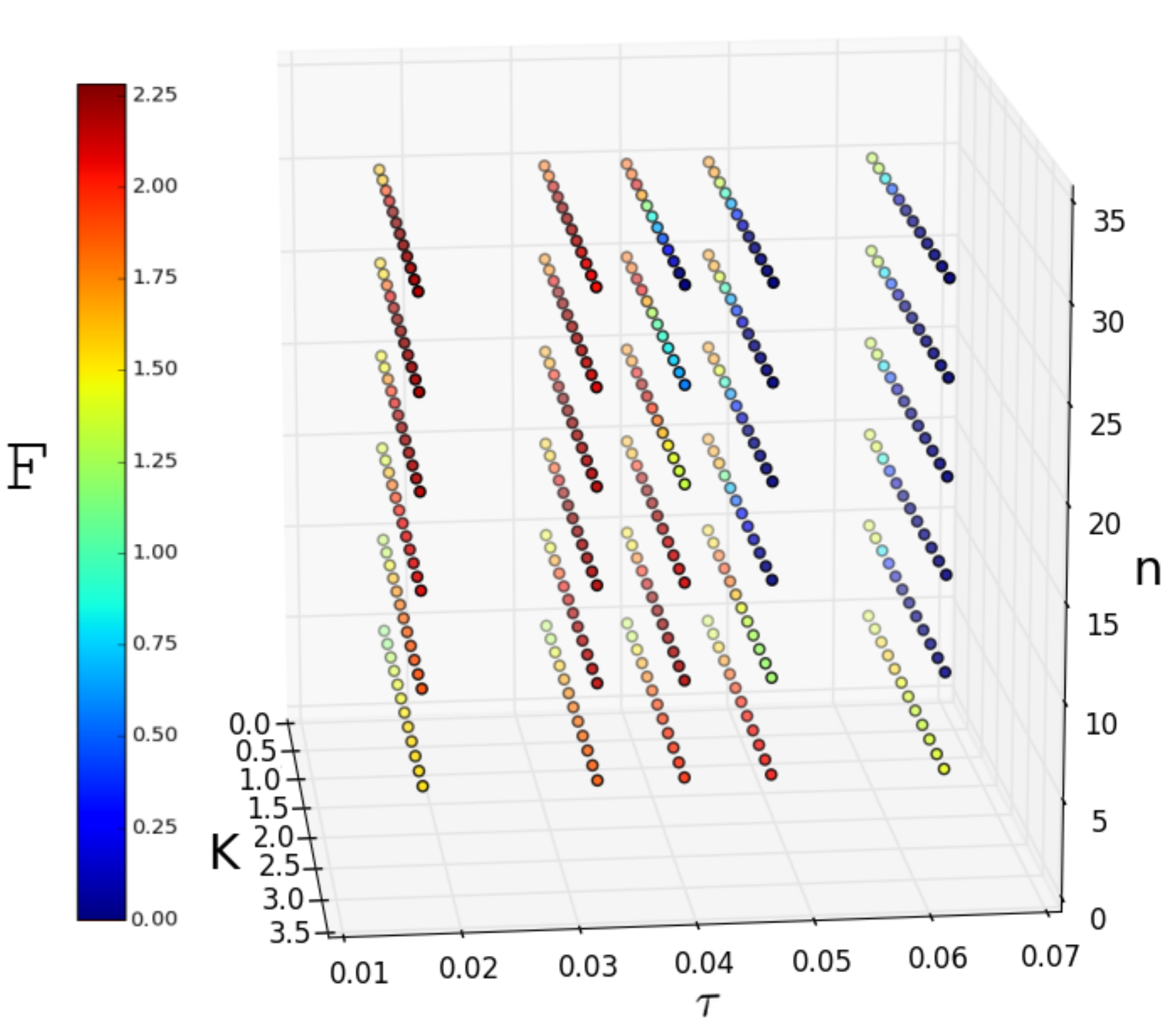}
\caption{The $F$-values applying to WEIN069 an intensity-driven gradient
  filter with parameters $K$, $\tau$, $n$ and $C_{x,y,z} = 5$. In this 
  3-D scatter plot, the $F$-values are displayed as a
  4-th dimension using a color scale. The red dots represents filtering 
  with an high value of the parameter $F$ ($F$-values $> 1.75$). The $F$-parameter 
  shows low values ($< 1$) for high values of $n$ and $\tau$
  ($n > 15$ and $\tau > 0.0475$).
  For more information regarding 
  the $F$-parameter refer to sections \ref{results} and \ref{noise}.
  }
\label{Ffinal}
\end{figure}

\noindent On the other hand, this filter applies a diffusion process 
which has the following drawbacks:
\begin{enumerate}[a)]
\item the flux scale is not conserved and depends on the signal-to-noise ratio and hence
      degree of \enquote*{smoothing} or resulting resolution;
\item setting too high values of the parameters $n$ and $\tau$ 
      can create unrealistic web structures (negative and positive) 
      between the peaks of the negative and positive parts of the noise. 
\end{enumerate}
The first issue is not a problem for visualization.
In fact, the main purpose of the filtering operation, in this context, 
is to find and enhance low-level signals. 
Quantitative analysis, such as calculating column densities,
intensity weighted mean velocities, velocity dispersions etc., 
can always be performed on the original data cube once the volume that 
contains all the signal has been identified.
Regarding the second issue: in Fig.~\ref{Ffinal} we show as a guideline
the dependencies of the $F$-parameter on the input parameters $K$, $\tau$ and $n$. 

Finally, the previous results suggest that intensity-driven gradient smoothing 
can be employed for finding H\,{\small I} sources as well.
This technique could be an alternative for the smooth-and-clip 
method and has the advantage that the user does 
not have to specify the smoothing kernels. The robustness of such a method should be tested on
a larger number of different cases than we have used here. 
This is beyond the scope of the present investigation.

In Section \ref{performance}, we reported the benchmark of our 
CPU and GPU implementations of the filtering algorithms investigated 
in this paper. The codes are publicly 
available\footnote{\url{https://github.com/Punzo/SlicerAstro/AstroSmoothing}}
and we integrated them in a module of $\tt{SlicerAstro}$\footnote{\url{http://wiki.slicer.org/slicerWiki/index.php/Documentation/Nightly/Extensions/SlicerAstro}}, 
a first design of an astronomical extension of $\tt{3DSlicer}$\footnote{ 
$\tt{3DSlicer}$ (\url{https://www.slicer.org/}) is a medical visualization package with advanced 
3-D visualization capabilities.} \citep{Slicer}. 
We showed that for data cubes with a number of voxels up to $5 \times 10^6$, 
GPU implementations of the smoothing filters can reach interactive
performance (maximum execution time, $T_p < 0.3\;s$)
exploiting a GTX860M,
i.e.\ a GPU suitable for gaming, 
found on laptops with mid-level performance.
For data cubes up to $10^8$ voxels,
the filters can still reach relatively  fast 
performance (maximum execution time with a GTX860M, $T_p < 3.5\;s$). 

In conclusion, the GPU implementation of the intensity-driven gradient filter
satisfies our filtering and visualization requirements best. 
The filter provides interactive performance, requires minimal 
tuning of the input parameters, and efficiently enhances faint structures 
in our data sample without degrading the resolution of the high
signal-to-noise data.

\section{Acknowledgments}
We thank M.A. Ramatsoku, E. Busekool and M.A.W. Verheijen 
for proving us the H\,{\small I} data of WEIN069 and NGC4111. 
We thank S. Pieper (Isomics) and K. Martin (Kitware)
of the 3D-Slicer project for their support with the 
GPU implementation of the filters.
We thank M. Papastergis and A. Marasco for their feedback regarding Section \ref{noise}.
Finally, we thank the reviewers for their constructive
comments, which helped us to improve the paper substantially.

3-D visualization screenshots shown in this paper were generated by using $\tt{3DSlicer}$.

D. Punzo and J.M van der Hulst acknowledge support from the European 
Research Council under the European Union's Seventh Framework Programme (FP/2007-2013)/ERC 
Grant Agreement nr. 291-531. 

\section{References}
\bibliographystyle{elsart-num-names}
\bibliography{Filtering.bib}

\end{document}